\documentclass[article]{aa}
\usepackage[utf8]{inputenc}
\usepackage[LGR,T1]{fontenc}
\usepackage{geometry}
\usepackage{xcolor}
\usepackage{subcaption}

\usepackage[left,modulo]{lineno}
\usepackage{setspace}

\usepackage{textcomp}
\usepackage{amstext}
\usepackage{graphicx}
\usepackage{upgreek}
\usepackage{txfonts}
\usepackage{amsmath}
\usepackage[]{hyperref}

\usepackage{lipsum, babel}

\hypersetup{colorlinks=true,citecolor=blue}

\definecolor{crimson}{rgb}{0.75, 0.0, 0.2}

\newcommand{\fig}[1]{Fig.\,\ref{#1}}
\newcommand{\figs}[2]{Figs.\,\ref{#1} and \ref{#2}}
\newcommand{\sect}[1]{Sect.\,\ref{#1}}

\def\sizefig{0.55}
\def\sizefigTp{0.65}
\def\sizefigHfix{0.25}

\def\mj{$M_\mathrm{J}$}
\def\rj{$R_\mathrm{J}$}
\def\ms{$M_\mathrm{S}$}
\def\rs{$R_\mathrm{S}$}
\def\he{He}
\def\co{CO}
\def\hho{H$_2$O}
\def\hh{H$_{2}$}
\def\tio{TiO}
\def\vo{VO}
\def\na{Na}
\def\k{K}
\def\Tp{$\nabla^+ T$ }
\def\Tm{$\nabla^- T$ }
\def\gcm{GCM }
\def\COratio{$[\mathrm{CO}]/[\mathrm{H}_2\mathrm{O}]$ }
\def\VOratio{$[\mathrm{CO}]/[\mathrm{VO}]$ }
\def\TiOratio{$[\mathrm{CO}]/[\mathrm{TiO}]$ }
\def\Heratio{$\frac{He}{H_2}$ }
\def\Piso{P_{\mathrm{iso}}}
\def\Tday{T_{\mathrm{d}}}
\def\Tnight{T_{\mathrm{n}}}
\def\alp{\alpha_{*}}
\def\taurex{TauREx }
\def\pyt{\textit{Pytmosph3R} }
\def\wasp{Wasp-121b }

\def\redchi{$\tilde{\chi}^2$}

\title{
Strong biases in retrieved atmospheric composition caused by day-night chemical heterogeneities}

\titlerunning{Strong biases in retrieved atmospheric composition caused by strong day-night chemical heterogeneities}

\author{William Pluriel\inst{1}
     \and
           Tiziano Zingales\inst{1}
           \and 
           J\'{e}r\'{e}my Leconte\inst{1}
           \and 
           Vivien Parmentier\inst{2}
}
           
\institute{Laboratoire d'astrophysique de Bordeaux, Univ. Bordeaux, CNRS, B18N, all\'{e}e Geoffroy Saint-Hilaire, 33615 Pessac, France
\and
Department of Physics, Oxford University, OX1 2JD, United Kingdom (vivien.parmentier@physics.ox.ac.uk)
}

\date{Received 2020 February 7; Accepted 2020 March 12}

\abstract{
   Most planets currently amenable to transit spectroscopy are close enough to their host star to exhibit a relatively strong day to night temperature gradient. For hot planets, this leads to cause a chemical composition dichotomy between the two hemispheres. In the extreme case of ultra hot jupiters, some species, such as molecular hydrogen and water, are strongly dissociated on the day-side while others, such as carbon monoxide, are not. However, most current retrieval algorithm rely on 1D forward models that are unable to model this effect. We thus investigate how the 3D structure of the atmosphere biases the abundances retrieved using commonly used algorithms. 
   We study the case of Wasp-121b as a prototypical ultra hot Jupiter. We use the simulations of this planet performed with the Substellar and Planetary Atmospheric Radiation and Circulation (SPARC/MIT) global climate model (GCM) and generate transmission spectra that fully account for the 3D structure of the atmosphere with \pyt. These spectra are then analyzed using the \taurex retrieval code. 
   We find that such ultra hot jupiter's transmission spectra exhibit muted \hho$\,$ features that originate in the night-side where the temperature, hence the scale-height, is smaller than on the day-side. However, the spectral features of molecules present on the day-side are boosted by both its high temperature and low mean molecular weight.
   As a result, the retrieved parameters are strongly biased compared to the ground truth. In particular the [CO]/[H$_2$O] is overestimated by one to three orders of magnitude. This must be kept in mind when using such retrieval analysis to infer the C/O ratio of a planet's atmosphere. We also discuss whether indicators can allow us to infer the 3D structure of an observed atmosphere. Finally we show that Wide Field Camera 3 from Hubble Space Telescope (HST/WFC3) transmission data of Wasp-121b are compatible with the day-night thermal and compositional dichotomy predicted by models. 
}

\begin{document}
\maketitle

\section{Introduction}

Since the discovery of the first exoplanets, observations have shown a great diversity of objects, from Earth-like planets to Ultra Hot Jupiters (UHJ).
Orbiting very close to their star, UHJs can reach an atmospheric temperature high enough to trigger the thermal dissociation of some of the chemical species, such as \hho$\,$ and \hh$\,$ \citep{Lodders2002, Visscher2006, Visscher2010}. They thus offer the opportunity to study the chemistry and physics of planetary atmospheres under extreme conditions, for which we have no equivalent in the Solar System. These interesting objects will be prime targets for new generation space missions like the James-Webb Space Telescope (JWST) \citep{Beichman2014} and Atmospheric Remote-sensing Infrared Exoplanet Large-survey (ARIEL) \citep{Tinetti2018}.

Currently, only a few UHJs have been discovered and studied using both transit and eclipse spectroscopy \citep{Wright2012, Haynes2015, Sheppard2017, Evans2017, Kreidberg_2018}. However, the analysis of UHJs is not simple, due to their complex chemical composition and dynamics.
Bayesian retrieval procedures being computationally-intensive, it is necessary to make strong assumptions to speed-up the atmospheric forward model. But a too simplistic set of assumptions can lead to strongly biased interpretations where the retrieved abundances are much higher than the expected ones.
\citet{Evans2017}, for example, used transit data to suggest presence of supersolar FeH abundance to fit HST-WFC3 data of \wasp and explained the reduced water shape around 1.3$\,\mu$m. \citet{Parmentier2018} later suggest that this could be due to the presence of partially CaTiO$_3$ cloudy atmosphere.

In the present study, we investigate whether the variations in composition inside the atmosphere of UHJs may affect transmission spectroscopy as severely as emission spectroscopy. Those tidally locked planets present a strong day/night contrast both in temperature \citep{Sudarsky_2000,Bell_2018,Arcangeli_2018} and in chemical heterogeneities due to the thermal dissociation of certain species such as \hho\ and \hh\ \citep{Parmentier2018,Marley2017}. As the thermal dissociation of the species is strongly linked to the temperature, the thermal day-night dichotomy entails a chemical dichotomy, with a day side devoid of water above about 100 mbar and a night side where water is present everywhere. However, other species such as CO requires higher temperature to dissociate \citep{Lodders2002} and are expected to remain constant in the atmosphere which would change the apparent [CO]/[H$_2$O] ratio. \citet{Caldas2019} developed a fully 3D model that allows to generate a transmission spectrum which considers the 3D structure of the atmosphere. They show that the light that goes through the planetary atmosphere carries the information of a portion of atmospheres that significantly extend around the limb. They highlight that the rays of light go through the day side first and then to the night side implying that strong 3D effects on the transmission spectrum across the limb occur when those effects along the limb are negligible. Hence, 1D transmission models cannot well take into account chemical heterogeneities describe above thus the need to use 3D transmission models.

\citet{Caldas2019} highlighted systematic biases on retrieved temperatures using a 1D retrieval model \taurex ~\citep{Waldmann2015a, Waldmann2015}. However, this earlier study only looked at atmospheres with a homogeneous composition to focus on thermal effects.
To that purpose, we focus on \wasp  \citep{Evans2016, Evans2017, Parmentier2018} as a prototype for UHJs. The complexity of the UHJ atmospheres describes in our \gcm model may suggest that we would need a more complex framework to analyze the transmission spectra of those particular planets. Thus, we will need fully 3D forward models to simulate realistic transmission spectra to better fit transit observations of UHJs. Hence, 1D retrieval models such as \taurex are probably biased in their analysis.

Hereafter, we first describe the observational chain used to simulate JWST observations in \sect{sec:observational chain}. Then, in \sect{sec:Modeling}, we explain the numerical experiments done to unravel the biases starting from a very simple, parametric modeling of \wasp to a more elaborate \gcm model. Moreover, we explain how thermal dissociation induces strong heterogeneities of composition between the day and night sides of the planet. We describe our retrieval results in \sect{sec:Results}. Finally, we highlight the main conclusion of our study and we discuss the limitations of our method and our models in \sect{sec:discus}.

\section{Presentation of our spectra generation and retrieval framework}
\label{sec:observational chain}

\subsection{SPARC/MITgcm global circulation model}
\label{sec:GCM}

We use the SPARC/MITgcm global circulation model~\citep{Showman2009} to model the atmosphere of WASP-121b. The model solves the primitive equations on a cubic-sphere grid. It has been successfully applied to a wide range of hot Jupiters~\citep{Showman2009,Kataria2015,Lewis2017,Parmentier2013,Parmentier2016} and has recently been used to study a few ultra-hot Jupiters in details~\citep{Kreidberg2018,Parmentier2018,Arcangeli2019}.

The model used here is exactly the same as described in~\citep{Parmentier2018}. Our pressure ranges from 200 bar to 2 $\mu$bar over 53 levels, We use a horizontal resolution of C32, equivalent to an approximate resolution of 128 cells in longitude and 64 in latitude. The radiative transfer is handled with a two-stream radiation scheme~\citep{Marley1999}, with the opacities treated using 8 correlated-k coefficients~\citep{Goody1989} withing each of our 11 wavelength bins~\citep{Kataria2013}.

Chemical equilibrium is assumed when calculating the opacities, meaning that thermal dissociation of molecules, including water and hydrogen is naturally taken into account. However, for practical purposes, we assume that the mean molecular weight and the heat capacity are constant throughout the atmosphere. The change in mean molecular weight is explored a-posteriori when projecting the GCM thermal structure in the \pyt grid (see Sect. \ref{pytmosph3R}). This model does not consider the impact of \hh\ recombination in the atmosphere which can have a non-negligible effect on the physics and the dynamics (\citealt{TK19}; see Sect. \ref{ssec: 3D Symmetric} for further details).

\begin{figure*}
\centering
\includegraphics[scale=\sizefig,trim = 1cm 1.3cm 2.7cm 0.cm, clip]{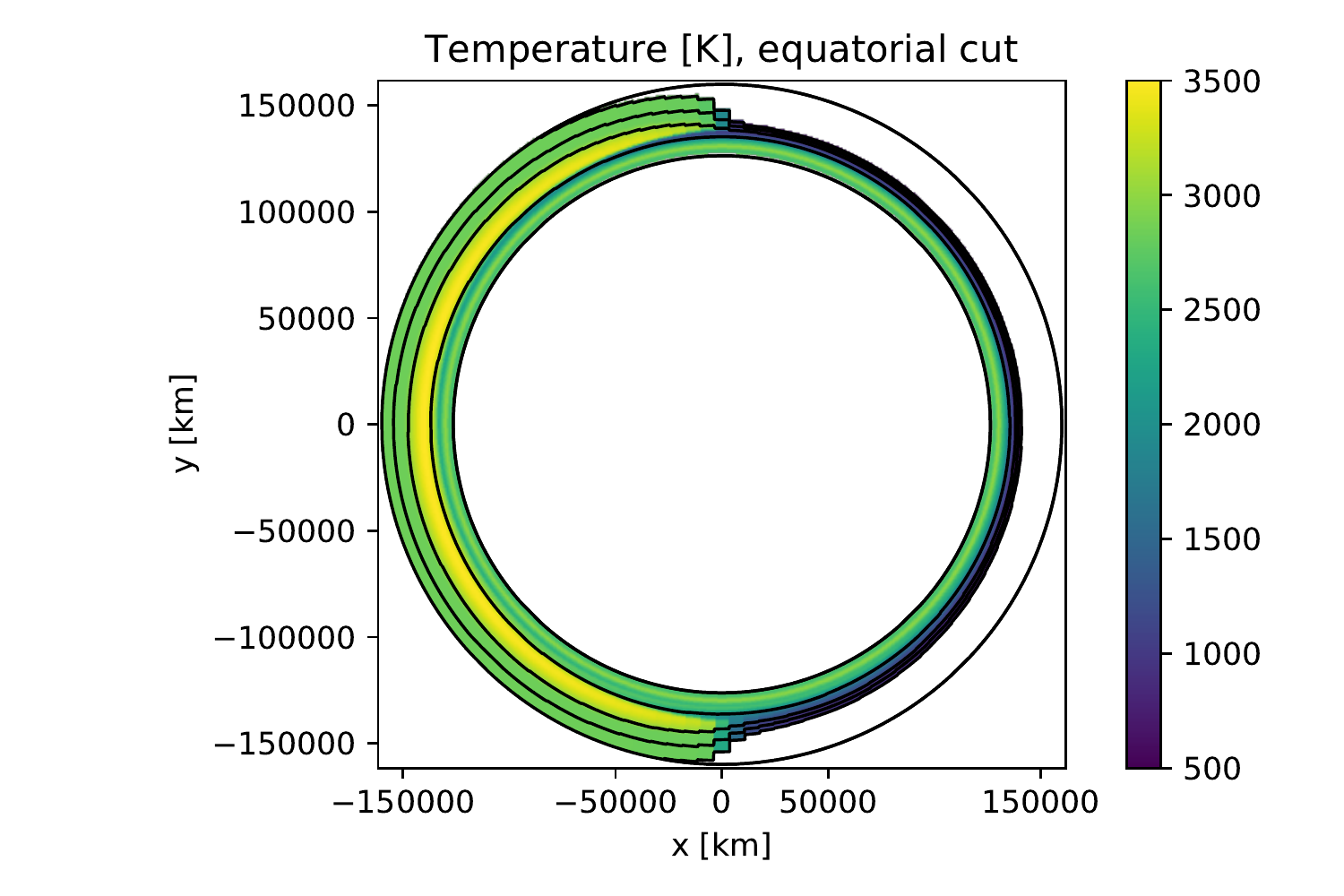}
\includegraphics[scale=\sizefig,trim = 4.1cm 1.3cm 2.7cm 0.cm, clip]{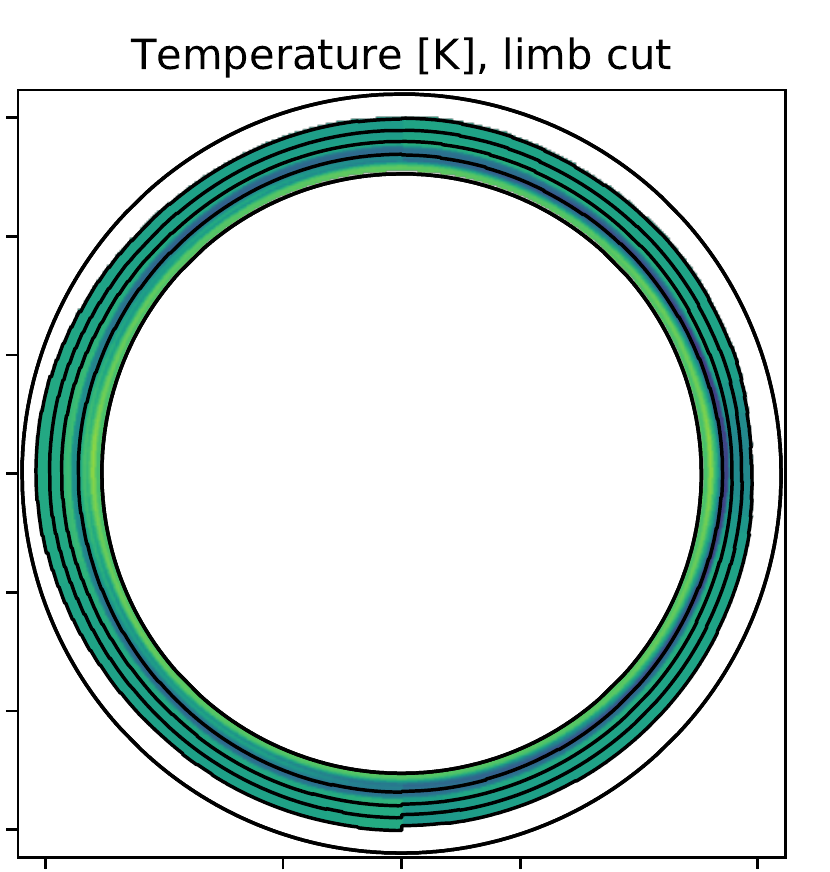}
\includegraphics[scale=\sizefig,trim = 4.1cm 1.3cm 0.cm 0.cm, clip]{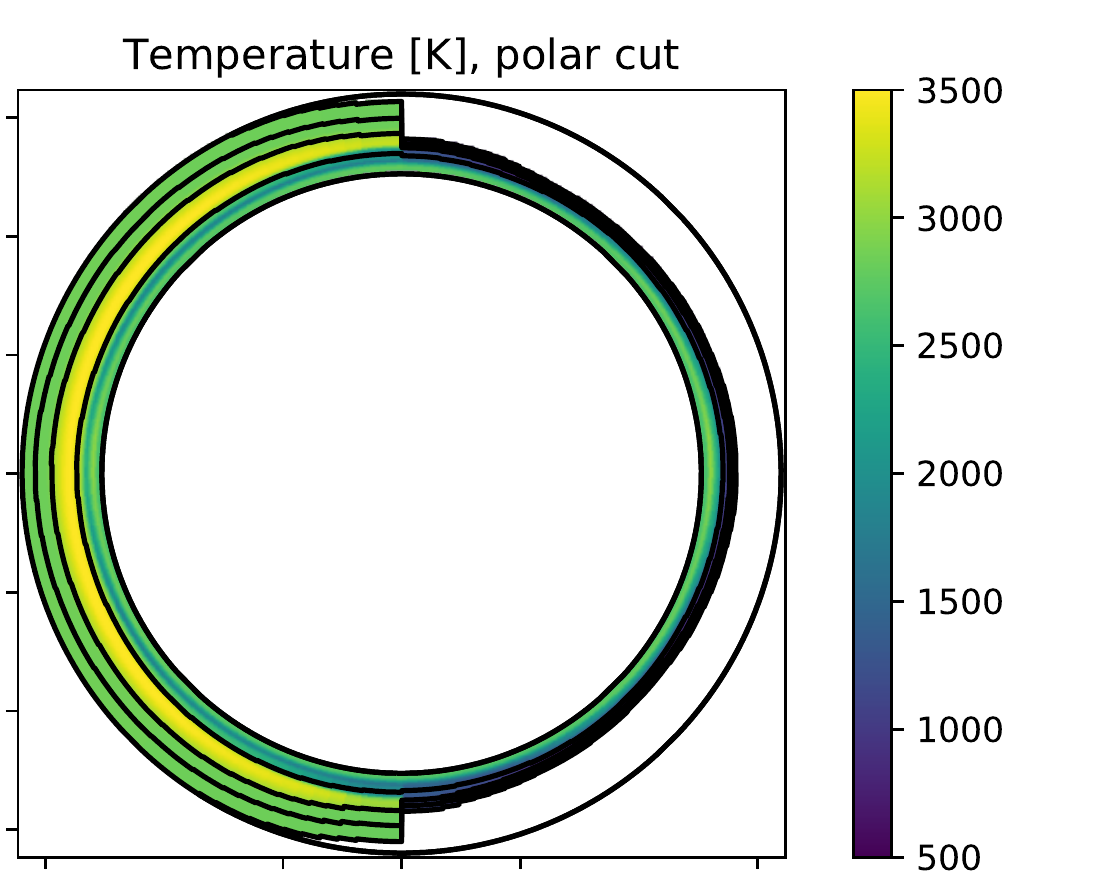}\\
\includegraphics[scale=\sizefig,trim = 1cm 0cm 2.7cm 0.cm, clip]{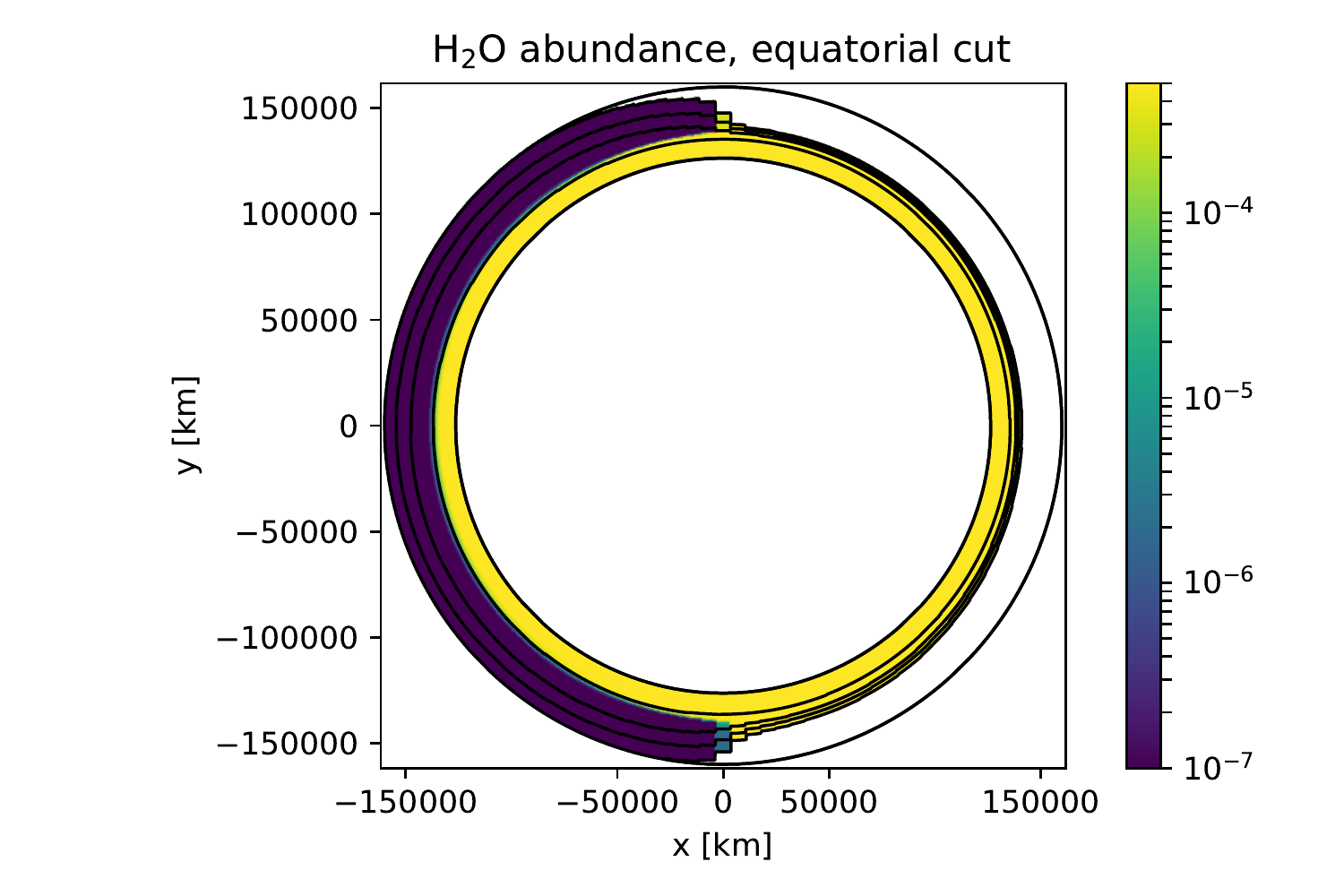}
\includegraphics[scale=\sizefig,trim = 4.1cm 0cm 2.7cm 0.cm, clip]{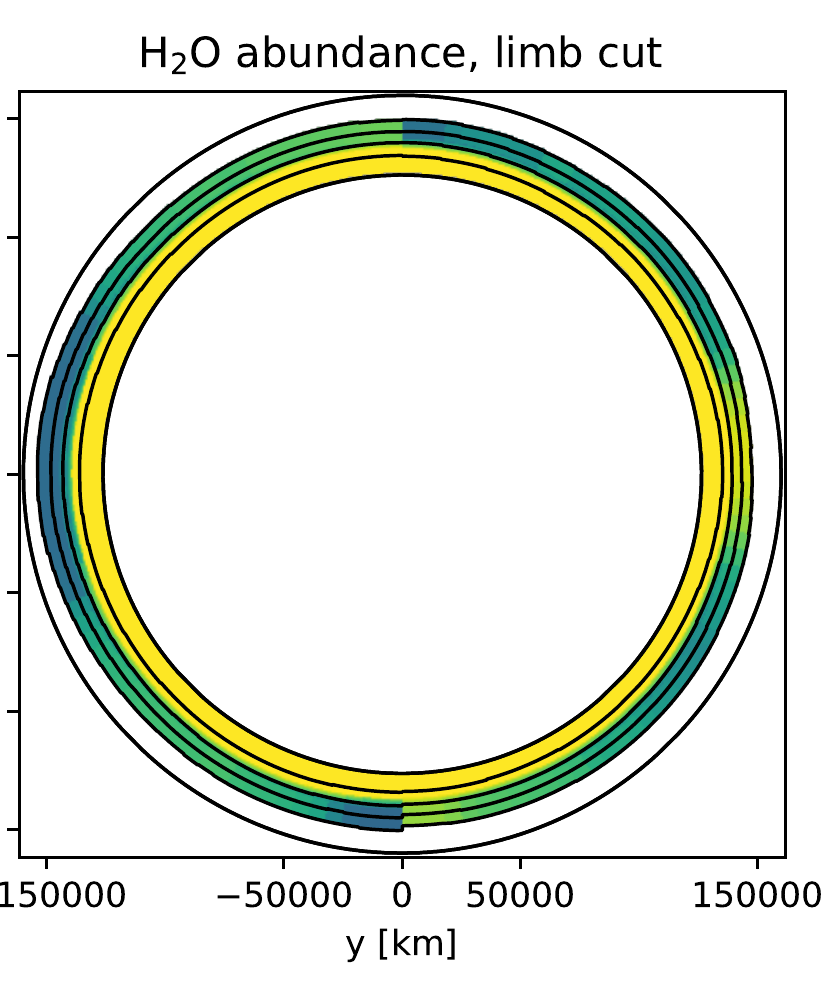}
\includegraphics[scale=\sizefig,trim = 4.1cm 0cm 0.cm 0.cm, clip]{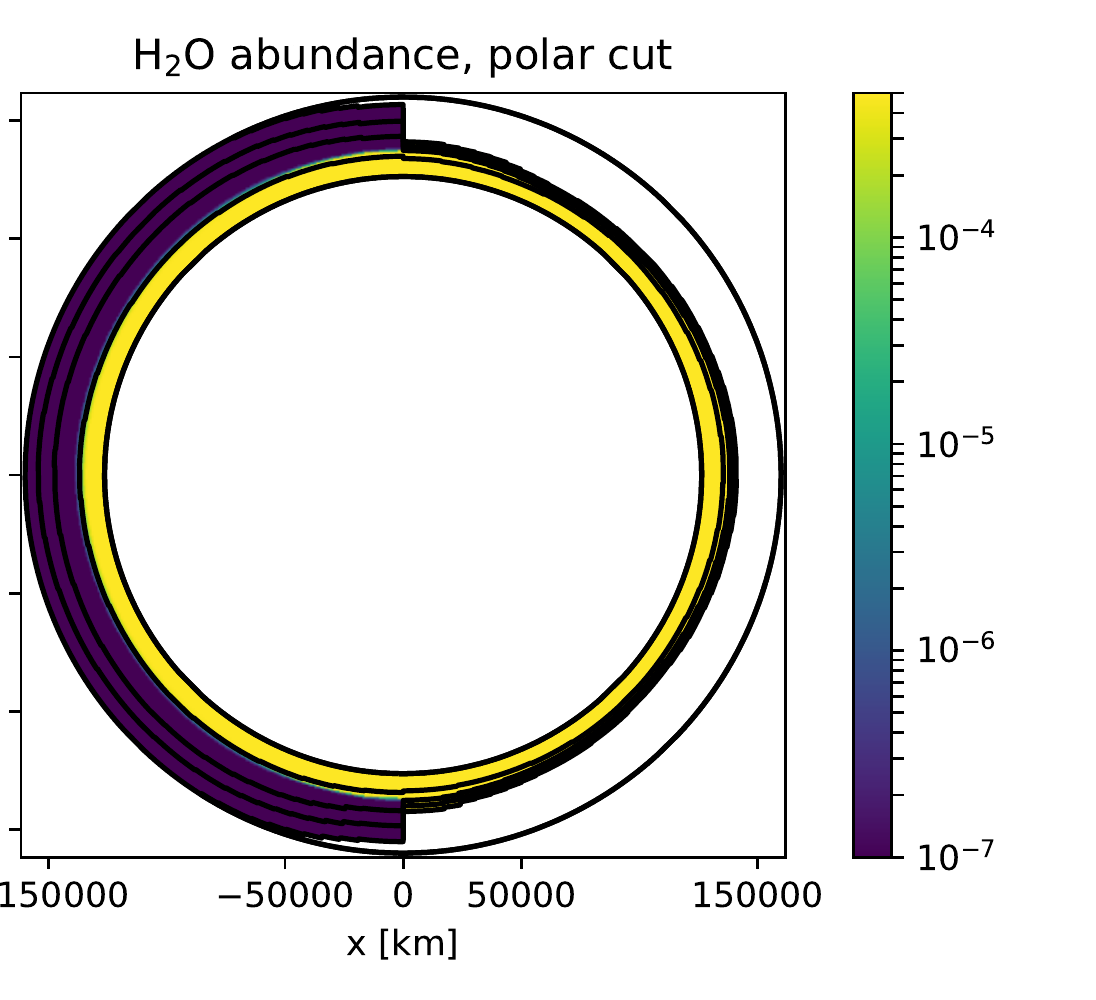}
\caption{GCM simulation of \wasp \citep{Parmentier2018} without \hh\ dissociation. Temperature (top) and the water abundance (bottom) for equatorial cut (left), limb cut (middle), and pole cut (right). From center outward, the 5 solid lines are respectively
the $1,434.10^7$, $10^3$, 1, $10^{-2}$, and $10^{-4}$ Pa pressure levels. The colormap for water abundance maps goes from $5.10^{-4}$ to $10^{-7}$. Note that the radius of the planet and the atmosphere are shown to scale.}
\label{fig: Atmospheric structure}%
\end{figure*}

\begin{figure*}
\centering
\includegraphics[scale=\sizefig,trim = 1cm 1.3cm 2.7cm 0.cm, clip]{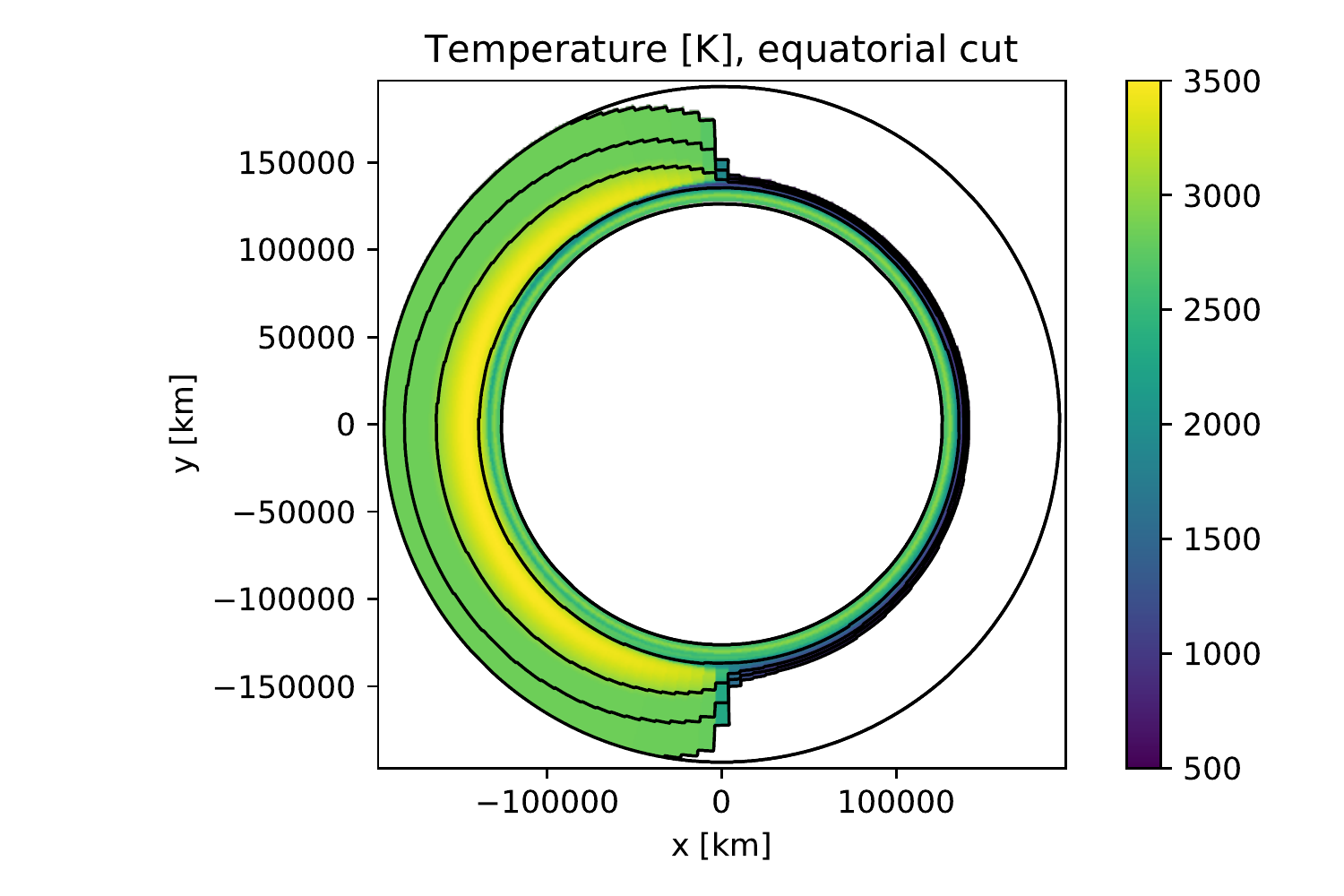}
\includegraphics[scale=\sizefig,trim = 4.1cm 1.3cm 2.7cm 0.cm, clip]{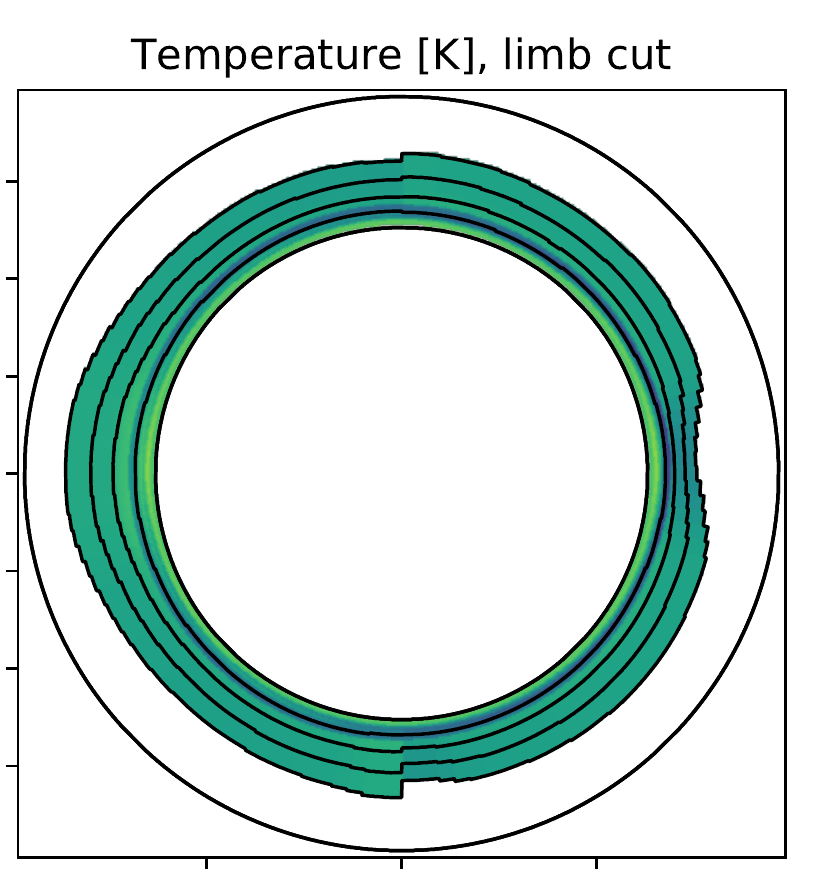}
\includegraphics[scale=\sizefig,trim = 4.1cm 1.3cm 0.cm 0.cm, clip]{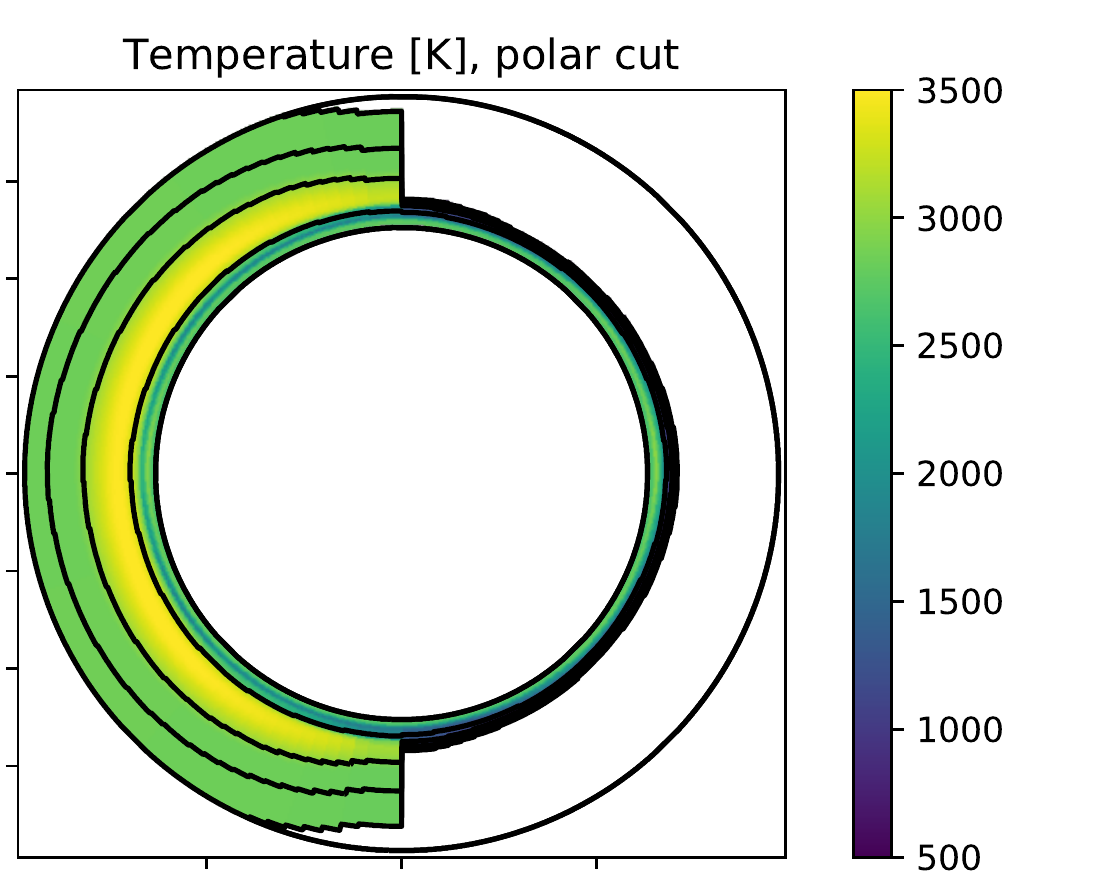}\\
\includegraphics[scale=\sizefig,trim = 1cm 0cm 2.7cm 0.cm, clip]{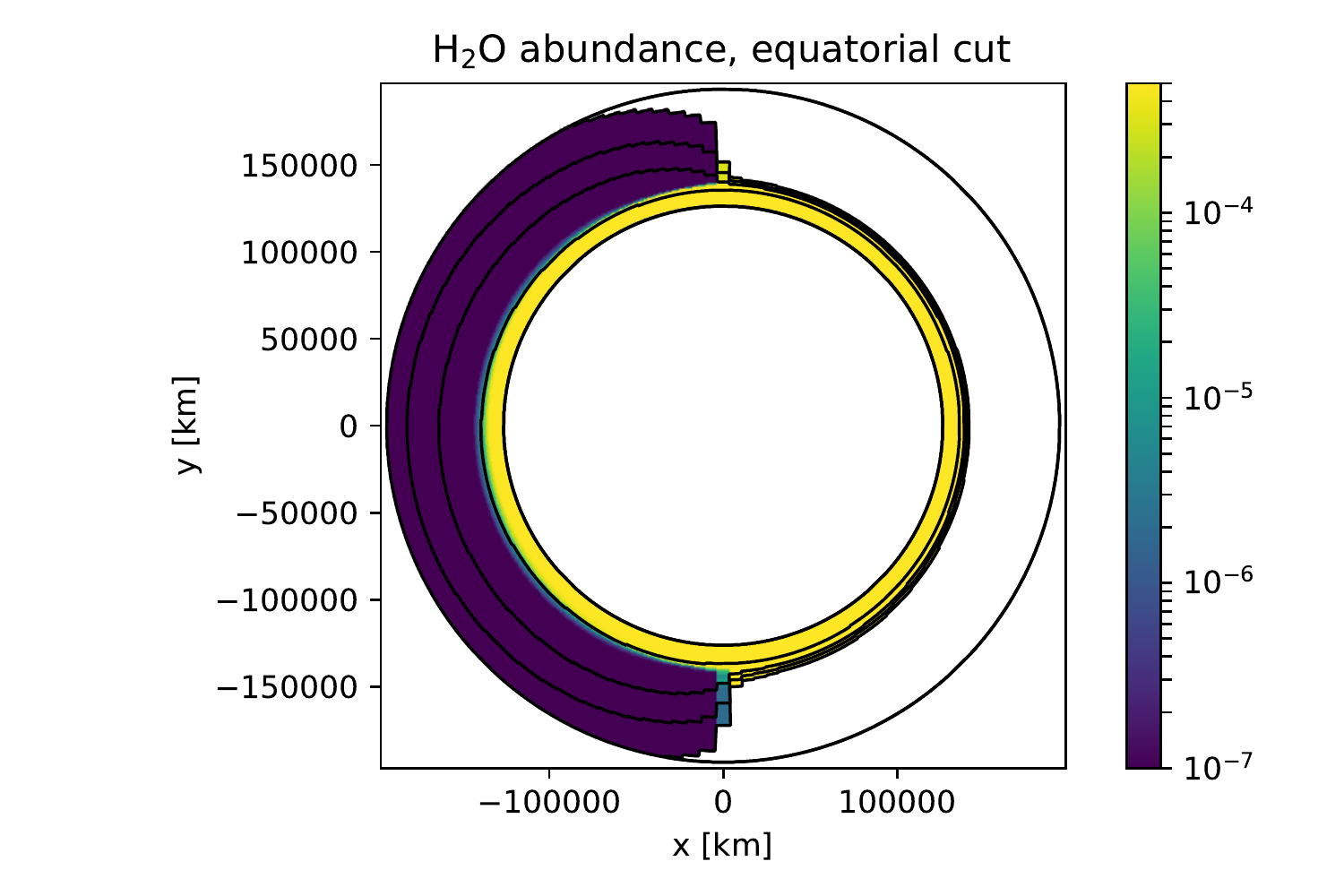}
\includegraphics[scale=\sizefig,trim = 4.1cm 0cm 2.7cm 0.cm, clip]{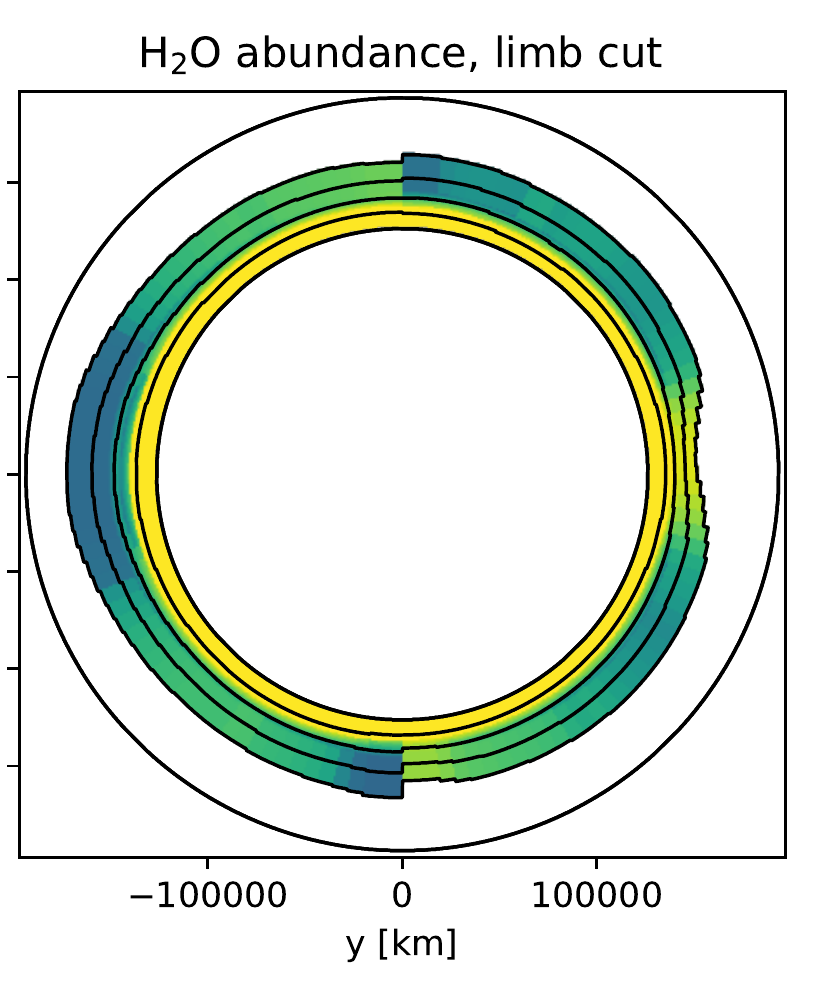}
\includegraphics[scale=\sizefig,trim = 4.1cm 0cm 0.cm 0.cm, clip]{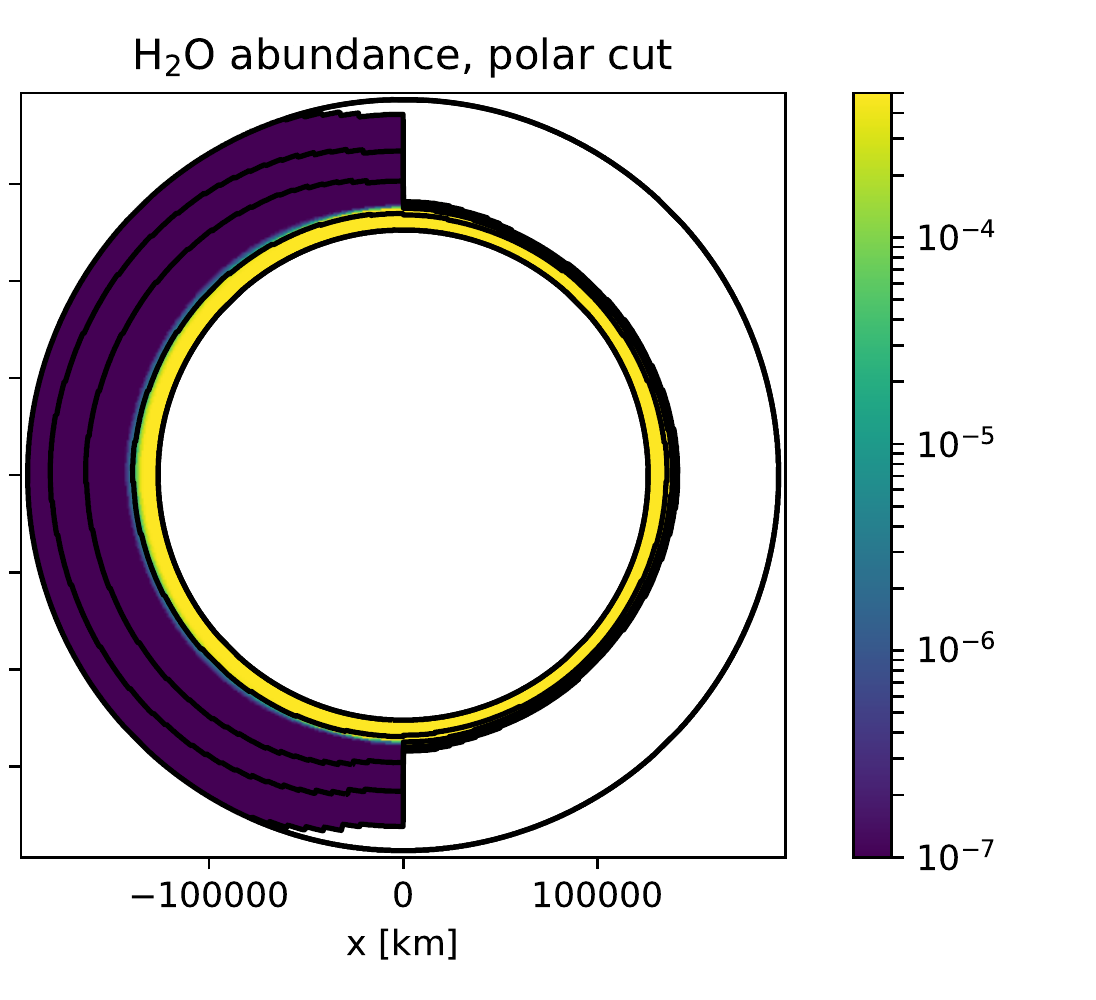}
\caption{Same plots as Fig \ref{fig: Atmospheric structure} but taking into account \hh dissociation. We clearly see that the dissociation of \hh\ mostly affects the day side of the atmosphere which increases even more the dichotomy between the day and the night side.}
\label{fig: Atmospheric structure H2 diss}%
\end{figure*}

Fig \ref{fig: Atmospheric structure}, \ref{fig: Atmospheric structure H2 diss} show the temperature maps generated by the \gcm which are structured in three main parts:
\begin{enumerate}
\item A quasi isothermal annulus around 2500 K from the surface pressure to approximately a pressure of 0.1 bar. We are here in the deep atmosphere, at high pressure, thus the redistribution of energy is very efficient due to jet stream.
\item Above this annulus, a cold night side where the temperature gradient decreases with altitude going from 1000 K to 500 K at whole latitudes.
\item Then, a very hot day side where the temperature gradient decreases with altitude going from 3500 K to 3000 K at whole latitudes. Those high temperatures enlarge the scale height implying a strong asymmetry in altitude between the day and the night side. 
\end{enumerate}
The transition between the day and the night side is very sharp, with a quasi isothermal terminator around 2200 K. We used in our study this global thermal structure to build idealize case of \wasp (\figs{fig: Atmospheric structure Tm}{fig: Atmospheric structure symcase}), as it will be explain in Sect \ref{sec:Modeling}.

As introduced before, the temperatures are high enough to allow dissociation of species, especially the water. We calculate the abundance of \hho\ in the \gcm simulations using the analytical equations provided by \citet{Parmentier2018} and we plot the abundances maps of \hho\ in Fig \ref{fig: Atmospheric structure}, \ref{fig: Atmospheric structure H2 diss}. We can see on the equatorial and polar maps that the water abundance is linked to the temperature since there is a total absence of water deep in the atmosphere on the day side, while the water abundance reaches solar abundance in the night side. The pressure dependence of water dissociation appears in the limb map, which is quasi-isothermal, since the water abundance reaches the solar abundance in the surface pressure then decreases to less than $10^{-12}$ at the top of the atmosphere (see Fig \ref{fig: Water_H2_abundances}). 

\subsection{Generation of transmission spectra with \pyt}
\label{pytmosph3R}

Transmission spectra are computed using \pyt\, as described in \citet{Caldas2019}. \pyt is designed to simulate transmission spectra based on any 3D atmospheric structure, including outputs of a Global Circulation Model (\gcm). It produces transmittance maps at any wavelength that can later be spatially integrated to yield the transmission spectrum.
The code can account for molecular opacities using the correlated-k method \citep{Fu1992} or monochromatic cross-sections calculated by ExoMol \citep{Yurchenko2011, Tennyson2012, Barton2013, Yurchenko2014, Barton2014}. In this study, we only use the latter to ensure a complete compatibility with the \taurex retrieval code \citep{Waldmann2015,Waldmann2015a}. Unless stated otherwise, our simulations only include \hh, \he, \hho,  and \co. We also take into account \hh-\hh$\,$ and \he-\hh$\,$ continua and the atmospheric scattering. To simplify the analysis, we did not include \tio\ and \vo\ to focus on the molecules above.

The abundances of the species in the atmosphere are calculated based on the temperature given by the \gcm simulation and following \citet{Parmentier2018} when dissociation occurs as described in Sect \ref{sec:GCM}. The molecular weight then derives from these abundances.
The gravity and the atmospheric scale height vary with the altitude from the planetary surface, i.e. the radius of the planet at a pressure of 10 bar, as described in \citet{Caldas2019}, and are defined, respectively, in Eq \eqref{gravity} and \eqref{scaleheight}:

\begin{equation}
g(z)=g_{0}\frac{1}{\left(1\,+\,\frac{z}{R_{p}}\right)^2}
\label{gravity}  
\end{equation} 

\begin{equation}
H(z)=\frac{kT(z)}{\mu(z)g(z)},
\label{scaleheight}  
\end{equation}

where $g_{0}$ is the surface gravity, $R_p$ is the planetary radius, $T$ is the temperature and $\mu$ is the mean molecular mass.

We define the volume mixing ratio (VMR) as:

\begin{equation}
x_\mathrm{i}=\frac{N_\mathrm{i}}{\sum{N_\mathrm{i}}},
\label{VMR_def}  
\end{equation}

where $N_\mathrm{i}$ is the molecular number density in molecules per volume unit. We also assume that all the others species are present in the atmosphere as trace gases, so we can write the number density of atomic Hydrogen:

\begin{equation}
N_\mathrm{H}=2\left(N_\mathrm{H_{2}}^{0}\,-\,N_\mathrm{H_{2}}\right)
\label{traces}  
\end{equation}

where $N_{H_{2}}^{0}$ is the molecular number density when no dissociation occur -- e.g. deep in the atmosphere (see Tab. ~\ref{tab:input_param}).

We ran \pyt at R=10000 resolution, with $R=\frac{\lambda}{\Delta\lambda}$, in the spectral range from 0.6$\,\mu$m to 10$\,\mu$m. 
The spectra are then binned down to a resolution of R=100 for the retrievals. We use this low spectral resolution for several reasons:

\begin{enumerate}
\item We simulate JWST observations from 0.6 to 10 $\mu$m using the low resolution prism mod provided by Near Infrared Spectrograph (NIRSpec) and Mid Infra-Red Instrument (MIRI) instruments \citep{Stevenson_2016}. This does not entail that using the prism configuration is the best observational strategy for such a bright target, but provides us with a uniformly sampled spectrum that does not arbitrarily put more weight in the retrieval on some spectral regions; 
\item We compared TauREx with the fully homogeneous case planet, as in \citet{Caldas2019}, and we saw that the optimum resolution for the best accuracy in the retrieval is the resolution R=100.
\item There are only two absorbing species in our atmosphere (H$_2$O and CO) and we do not need to resolve specific lines, but large features in large spectral bands;
\end{enumerate}

In all our simulations we used the planetary and stellar parameters shown in Tab ~\ref{tab:wasp121_param}. In the part of the atmosphere where \hh\ does not dissociate, we have a fixed \Heratio ratio at 0.25885. Then, we compute the volume mixing ratios of H and He by combining the equations \eqref{VMR_def} and \eqref{traces} where \hh\ dissociates.

Since the overall flux is dominated by the star, we assume that the spectral noise is dominated by the stellar photon noise, defined as:

\begin{equation}
N_\mathrm{phot} = \frac{\pi \tau \Delta t}{hc} \left( \frac{R_*D}{2d}\right)^2 \int_{\lambda_1}^{\lambda_2} B(\lambda, T_*)\lambda d \lambda,
\label{photon_noise}
\end{equation}
where $\lambda_1$ and $\lambda_2$ are the limiting wavelengths of the bin considered, $d$ is the distance of the star (270 pc for WASP-121) and $R_*$, $T_*$ are, respectively, the stellar radius and the effective temperature. $D$, $\tau$ and $\Delta t$ are, respectively, the telescope diameter, the system throughput and the integration time, whose values has been fixed for JWST according to \citet{2015PASP..127..311C}.

Using Eq \ref{photon_noise}, the uncertainty varies from 10ppm to 50ppm between 1 and 10$\mu$m, assuming a single WASP-121b transit.
Since systematics may prevent us from reaching a 10ppm precision with JWST, wherever the shot noise was lower than 30ppm we assumed a floor noise of 30ppms through the whole spectral domain \citep{Greene_2016}.

Note that we compute different 3D structures as input for \pyt with the parameters described in table ~\ref{tab:input_param}. We will describe in details those structures in section \ref{sec:Modeling}.

\begin{table*}
\hfill{}%
\begin{tabular}{|c|c|c|c|c|c|}
\hline 
Input parameters & Min-Max temperature [K] & angle $\beta$ & $P_\mathrm{bot}$ [Pa] & $P_\mathrm{top}$ [Pa]  & $x_\mathrm{CO}$ [log(VMR)] \tabularnewline
\hline 
Symmetric case \Tm & 1400-2800 & 20° & $10^{6}$ & $10^{-4}$ & -3.36 \tabularnewline
\hline 
Symmetric case \Tp & 500-3500 & 10°& $10^{6}$ & $10^{-4}$ & -3.36 \tabularnewline
\hline 
\gcm case & 493-3545 & - & $1.434\times10^{7}$ & $10^{-4}$ & -3.36 \tabularnewline
\hline 
\end{tabular}\hfill{}
\caption{Parameters used in the three simulation configurations of WASP-121b. \label{tab:input_param}}
\end{table*}

\begin{table*}
\begin{center}
\begin{tabular}{|c|c|c|c|c|}
\hline 
Star & Mass [\ms] & Radius [\rs] & Temperature [K] & Bibliographical references \tabularnewline
\hline 
Wasp-121 & 1.353 & 1.458 & 6460 &  \citet{Delrez2016}
\tabularnewline
\hline 
\end{tabular}
\\
\begin{tabular}{|c|c|c|c|c|}
\hline
Planet & Mass [\mj] & Radius [\rj] & Temperature [K] & Bibliographical references \tabularnewline
\hline 
Wasp-121 b & 1.183 & 1.8 & 2360 & \citet{Delrez2016}, \citet{Evans2016} \tabularnewline
\hline 
\end{tabular}
\caption{Mass, radius and temperature parameters of the WASP-121 star (top) and its planet Wasp-121 b (bottom). They are used both for the observable generations and the retrievals. \label{tab:wasp121_param}}
\end{center}
\end{table*}

\subsection{Retrieval model: \taurex}
\label{TauREx}

The spectral retrievals has been computed with the \taurex retrieval code \citep{Waldmann2015,Waldmann2015a}. \taurex consists in a Bayesian retrieval code which uses the ExoMol line lists \citep{Yurchenko2011, Tennyson2012, Barton2013, Yurchenko2014, Barton2014}, HITRAN \citep{Rothman2009, Gordon2013} and HITEMP \citep{Gordon2010}.
\taurex assumes a plane parallel exoplanetary atmosphere with a one-dimensional distribution elements and temperature-pressure profile. We assume the radius at the base of the model of \wasp as the value for which the atmospheric pressure is 10 bar. We simulate the exoplanetary atmosphere assuming a pressure between $10^6$ and $10^{-4}$ Pascals. We consider an atmosphere dominated by Hydrogen and Helium, with a mean molecular weight of 2.3 amu. We also suppose the presence of \co\ and \hho\ as trace gases.

An important assumption is that \taurex does not take into account thermal dissociation of species and assumes the molecular abundances constant with the atmospheric pressure (see \citet{Changeat_2019} for a detailed discussion on the effect of this assumption). We set up the prior range for the molecular volume mixing ratios of \co\ and \hho\ between $10^{-12}$ and $10^{-1}$ which allow us to explore a large range of solutions. Even if we never add clouds in our input simulation, we set up the prior range for clouds between $10^{-12}$ and $10^{-1}$ to allow the retrieval code to have more freedom. We also feel that this better simulates a real situation where clouds would be put in the retrieval without prior knowledge about their occurrence in the real planet. The allowed range for the planetary radius and the atmospheric temperature (assumed constant in the vertical) are, respectively, $1.3R_{Jup}-4.0R_{Jup}$ and $400K-3600K$, with $R_{Jup}$ the radius of Jupiter. The range for the temperature is linked to the minimal and maximal temperature in the \gcm simulation of \wasp$\,$ as shown in Tab \ref{tab:input_param}. As we simulate JWST observations, we set up the range of wavelength between 0.6 to 10 $\mu$m. For the non-retrieved parameters such as masses and stellar radius, we use those describe in Tab \ref{tab:wasp121_param}.

We note that the \taurex retrieval code is not designed to take into account the thermal dissociation of \hh. When \hh\ thermal dissociation occurs, the \he/\hh\ ratio changes as a function of pressure and temperature. The mean atmospheric weight can also be lower than 2amu. In order to be consistent across all our sets of simulations, we used the same configuration file for our spectral retrievals and we left the \he/\hh\ ratio constant. By doing so we could estimate whether an atmosphere where \hh\ dissociation does not occur can explain our input spectrum or not. We will come back to this in the Sect \ref{h2ovar-h2var}.

Again, some of these assumptions may seem too simplistic in view of the physics we envision in the real atmospheres of UHJs, but we think that it is important to understand how such simplistic assumptions -- commonly used nowadays -- can bias the conclusions from a retrieval analysis. 

\section{Input models: from simple case to GCM simulations}
\label{sec:Modeling}

In order to understand and to characterize the biases due to the atmospheric 3D structure of \wasp atmosphere, we decided to start from a simple case, and add, as the study progresses, further levels of complexity. Using this approach, we will show how to distinguish the observed biases. Note that we used the same atmospheric composition in all our cases, which is \he, \hh, \hho\ and \co.

The validation of \pyt's model with \taurex was done by \citet{Caldas2019}. So, we started by considering the atmospheric 3D structure of an idealized \wasp. In Figs \ref{fig: Atmospheric structure Tm} and \ref{fig: Atmospheric structure symcase} we show the 3D temperature maps used by \pyt to calculate the transmission spectrum. With this model we neglect effects which depend on the altitude, using isothermal pressure-temperature profiles in each cell, and we focus on the compositional effects.
Finally, we simulated a 3D atmospheric structure of \wasp using a Global Circulation Model (GCM) which takes into account both vertical and horizontal effects.

\subsection{Idealized 3D atmosphere with day to night gradient}
\label{ssec: 3D Symmetric}

\subsubsection{Effects of the day-night temperature gradient}

First, we create an idealized case of \wasp atmospheric structure that is symmetric around the planet/observer line of sight. The main parameters to simulate this simple structure are $T_\mathrm{d}$ and $T_\mathrm{n}$ and $\beta$, respectively the day-side, the night-side and the opening angle. $T_\mathrm{d}$ and $T_\mathrm{n}$ are correlated respectively to the sub-stellar and the anti-stellar point. The opening angle $\beta$ describes how the transitions between the day and the night side is computed, the smaller the sharper. We constructed our model by going symmetrically from the day side to the night side temperature considering a linear transition between the two. To avoid as much as possible effects due to vertical temperature gradients, every column of our model follows an isothermal profile above a given pressure level ($\Piso=0.13\,\mathrm{bar}$). In this part of the atmosphere, the temperature however varies with respect to the local solar elevation angle, $\alp$. In the deeper parts of the model, in accordance with results from a GCM  for \wasp (see \sect{sec:GCM}), the temperature is uniform and set to 2500\,K. In summary, for a given pressure and local solar elevation angle, the temperature is given by:

\begin{equation} 
\left\{ \begin{array}{rl} 
P>\Piso & T = \Tday \\ 
P<\Piso & \left\{ \begin{array}{rl} 
2\alp>\beta & T = \Tday \\ 
\beta>2\alp>-\beta & T = \Tnight+(\Tday-\Tnight)\frac{\alp + \frac{\beta}{2}}{\beta} \\
2\alp<-\beta & T = \Tnight
\end{array} \right.
\end{array} \right.
\end{equation}

For \wasp we considered two end-member cases:

\begin{enumerate}
 \item The high gradient model (\Tp), which goes from  $500\,\mathrm{K}$ to $3500\,\mathrm{K}$ using $\beta=10^\circ$. This one is supposed to be representative of the day-night gradient present in the GCM without the heterogeneities in the vertical direction and along the limb.
 \item The low gradient model (\Tm), which goes from  $1400\,\mathrm{K}$ to $2800\,\mathrm{K}$ using $\beta=20^\circ$. This case is more representative of GCM simulations accounting for the thermal effect of \hh recombination. We anticipate this case to be conservative and to provide a lower bound for the real bias caused by day-night temperature differences.
\end{enumerate}

\begin{figure*}
\centering
\includegraphics[scale=\sizefigTp,trim = 1.5cm 1.3cm 0.cm 0.cm, clip]{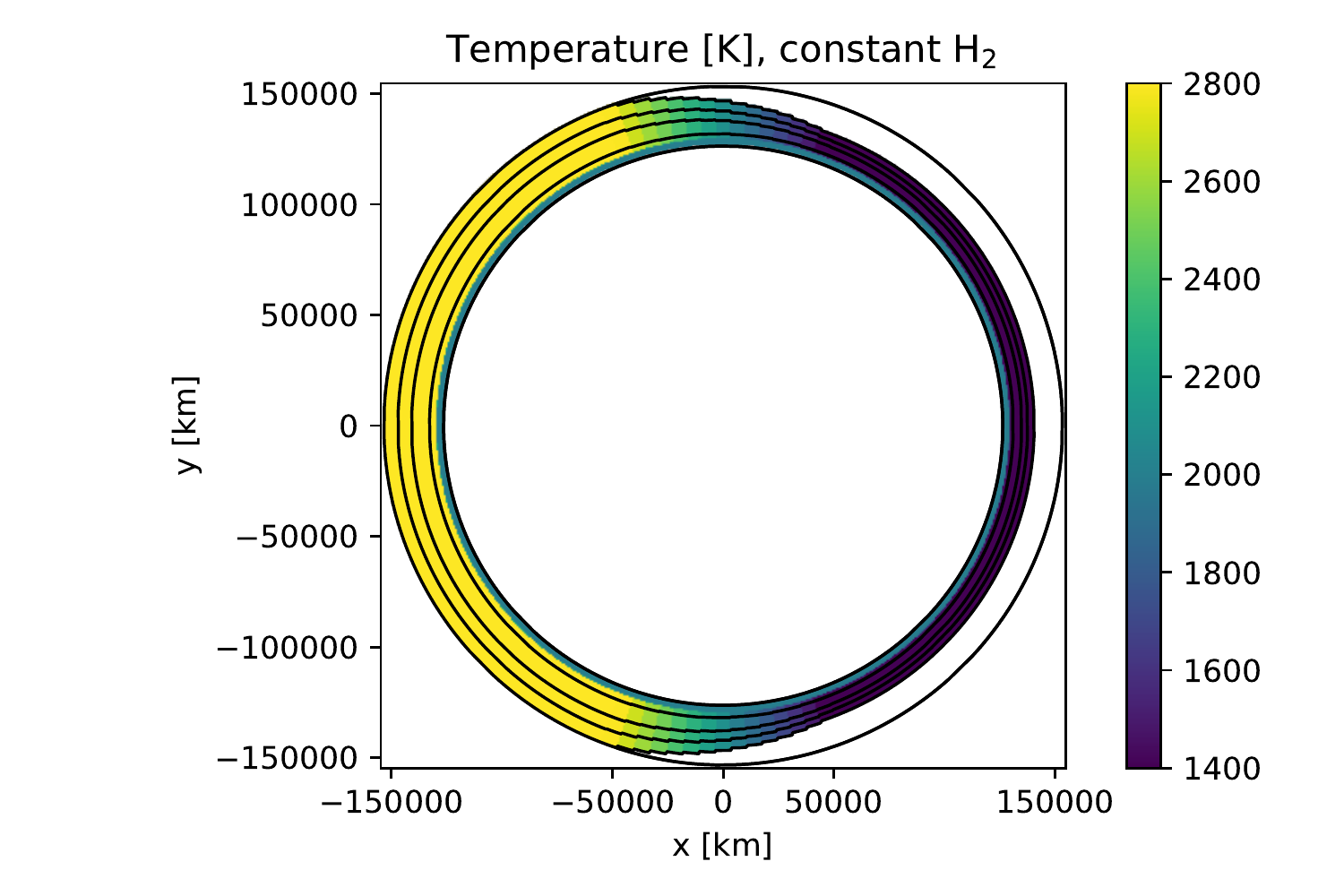}
\includegraphics[scale=\sizefigTp,trim = 1.5cm 1.3cm 0.cm 0.cm, clip]{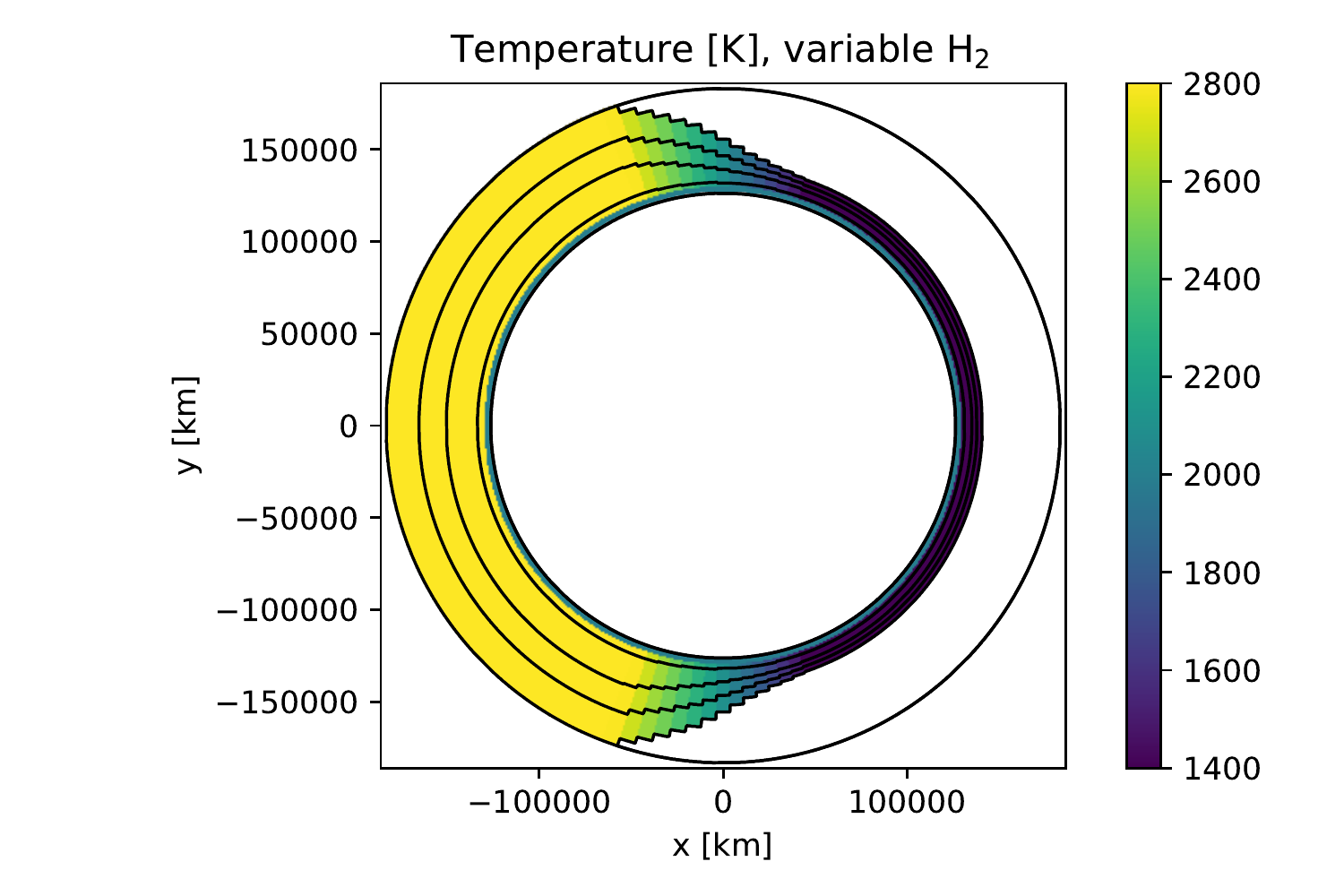}\\
\includegraphics[scale=\sizefigTp,trim = 1.5cm 0cm 0.cm 0.cm, clip]{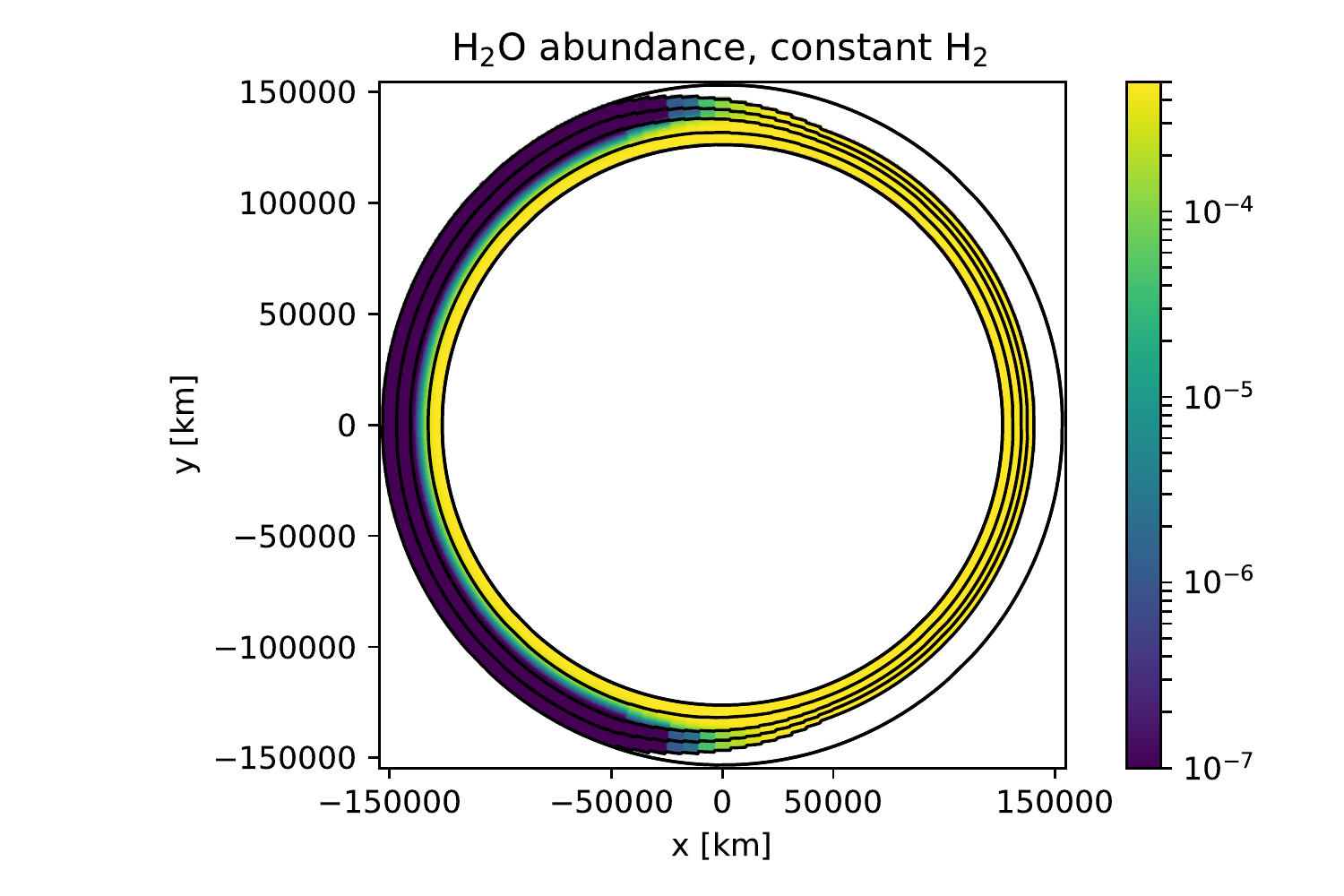}
\includegraphics[scale=\sizefigTp,trim = 1.5cm 0cm 0.cm 0.cm, clip]{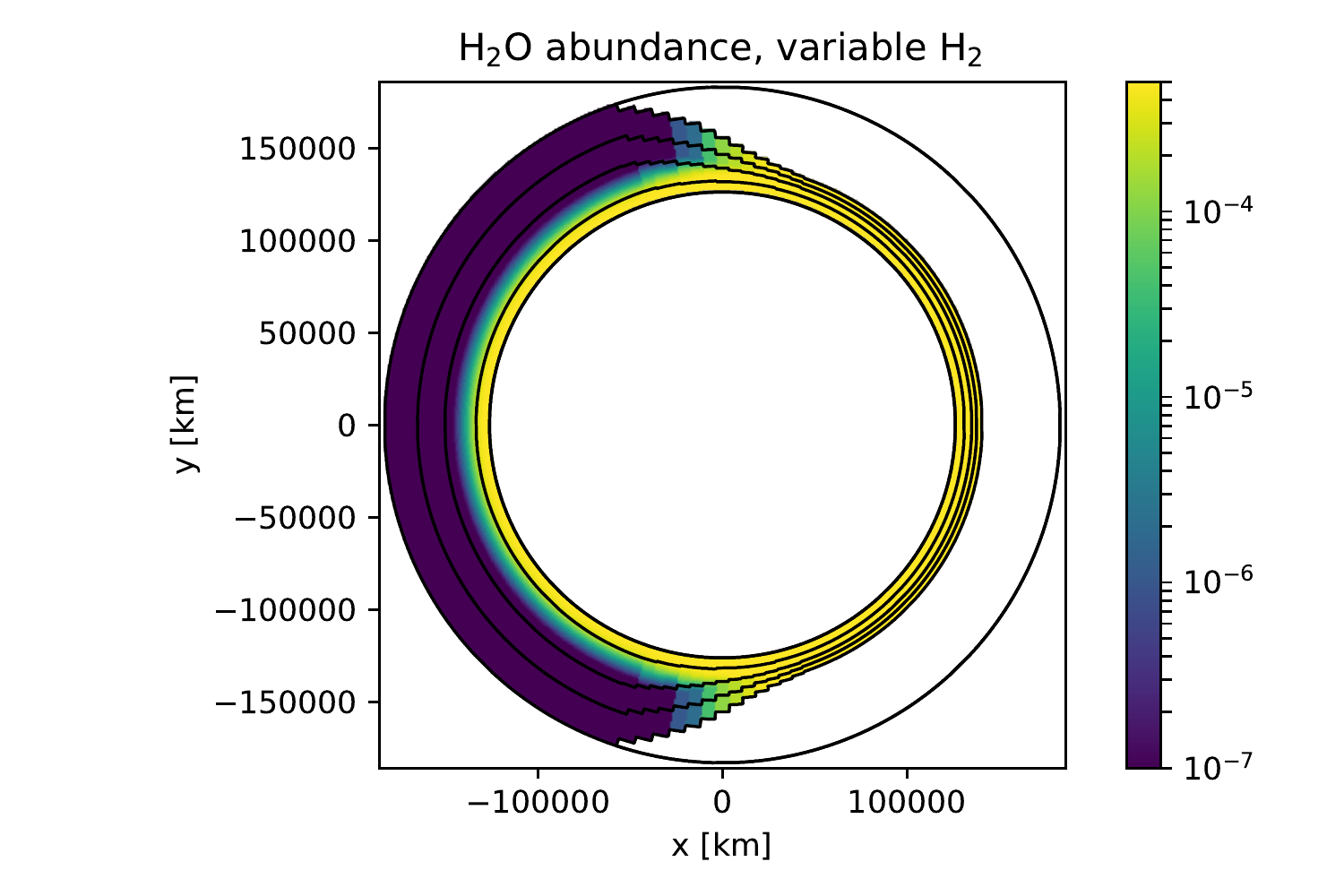}
\caption{Atmospheric structure of the symmetric, idealized case \Tm, assuming an absence and the presence of \hh\ dissociation (left and right respectively). We show the temperature map (top) and the water abundance map (bottom). The temperature gradient goes from $\mathrm{1400\,K}$ to $\mathrm{2800\,K}$. The transmission angle is 20° around the terminator line and have a $\mathrm{2500\,K}$ ring at P=$\mathrm{0.13\,bar}$ from the surface pressure. From center outward, the 5 solid lines are, respectively, the $10^6$, $10^3$, 1, $10^{-2}$, and $10^{-4}$ Pa pressure levels. The colormap for water abundance maps goes from $5.10^{-4}$ to $10^{-7}$. Note that the radius of the planet and the atmosphere are shown to scale.}
\label{fig: Atmospheric structure Tm}%
\end{figure*}

\begin{figure*}
\centering
\includegraphics[scale=\sizefigTp,trim = 1.5cm 1.3cm 0.cm 0.cm, clip]{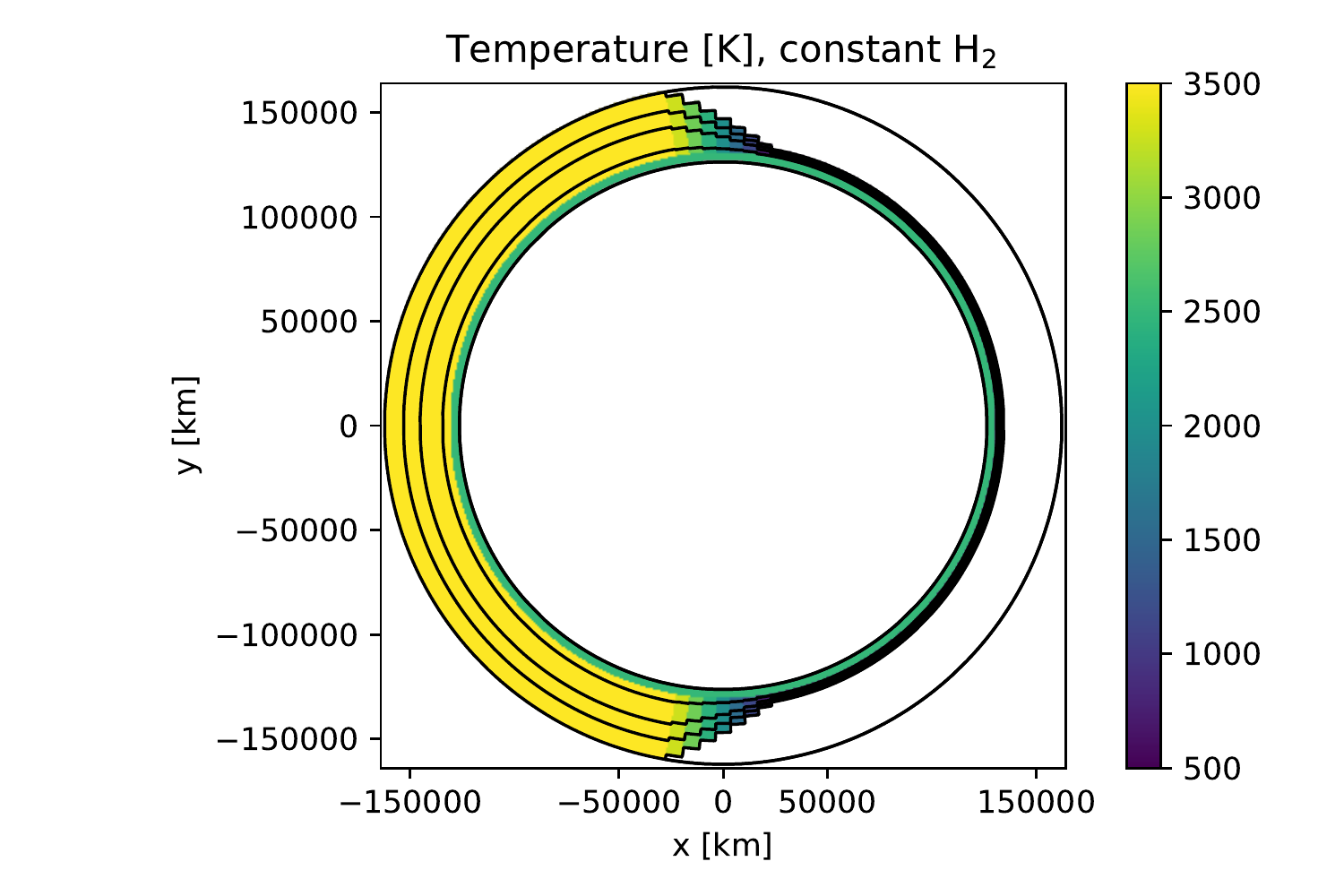}
\includegraphics[scale=\sizefigTp,trim = 1.5cm 1.3cm 0.cm 0.cm, clip]{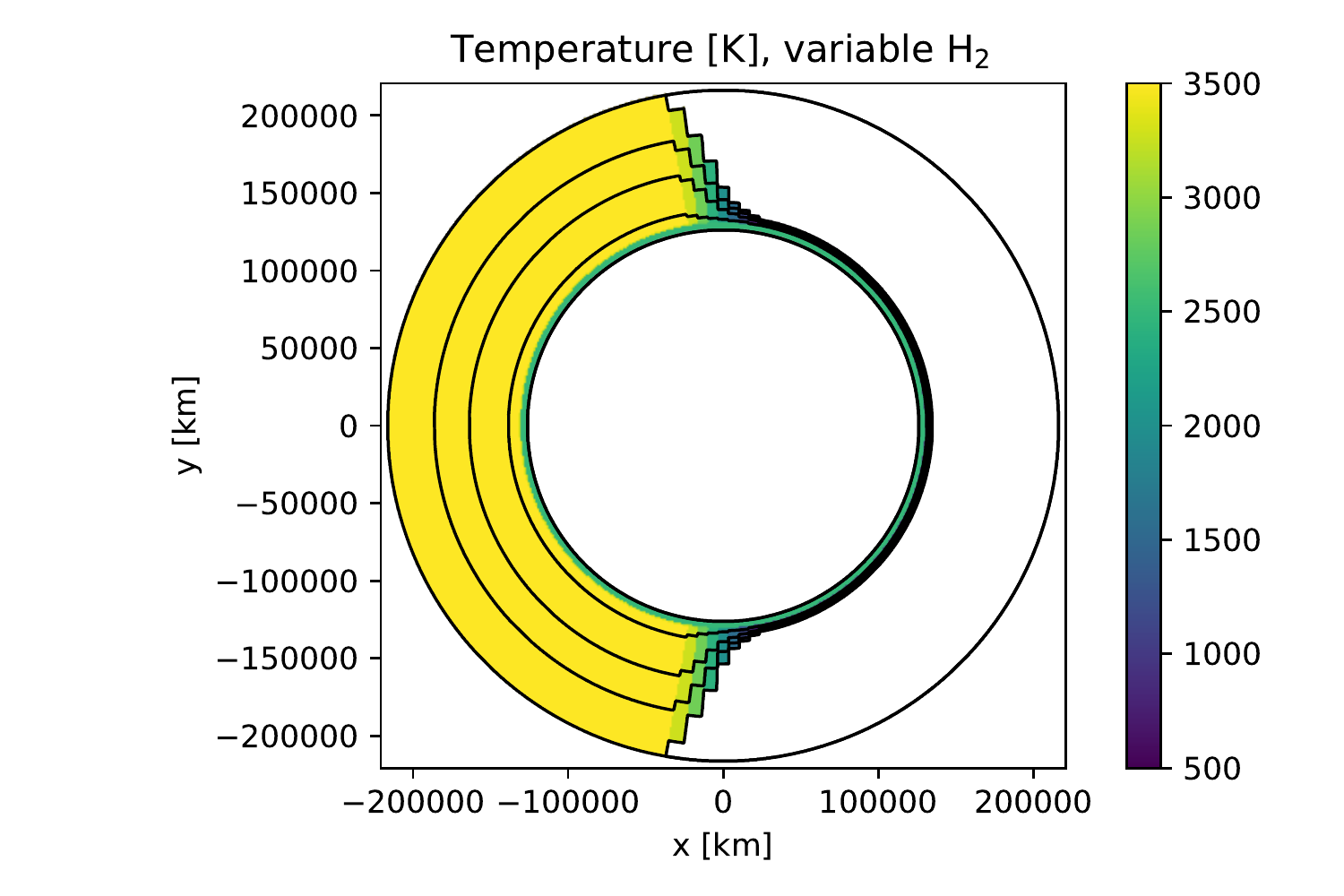}\\
\includegraphics[scale=\sizefigTp,trim = 1.5cm 0cm 0.cm 0.cm, clip]{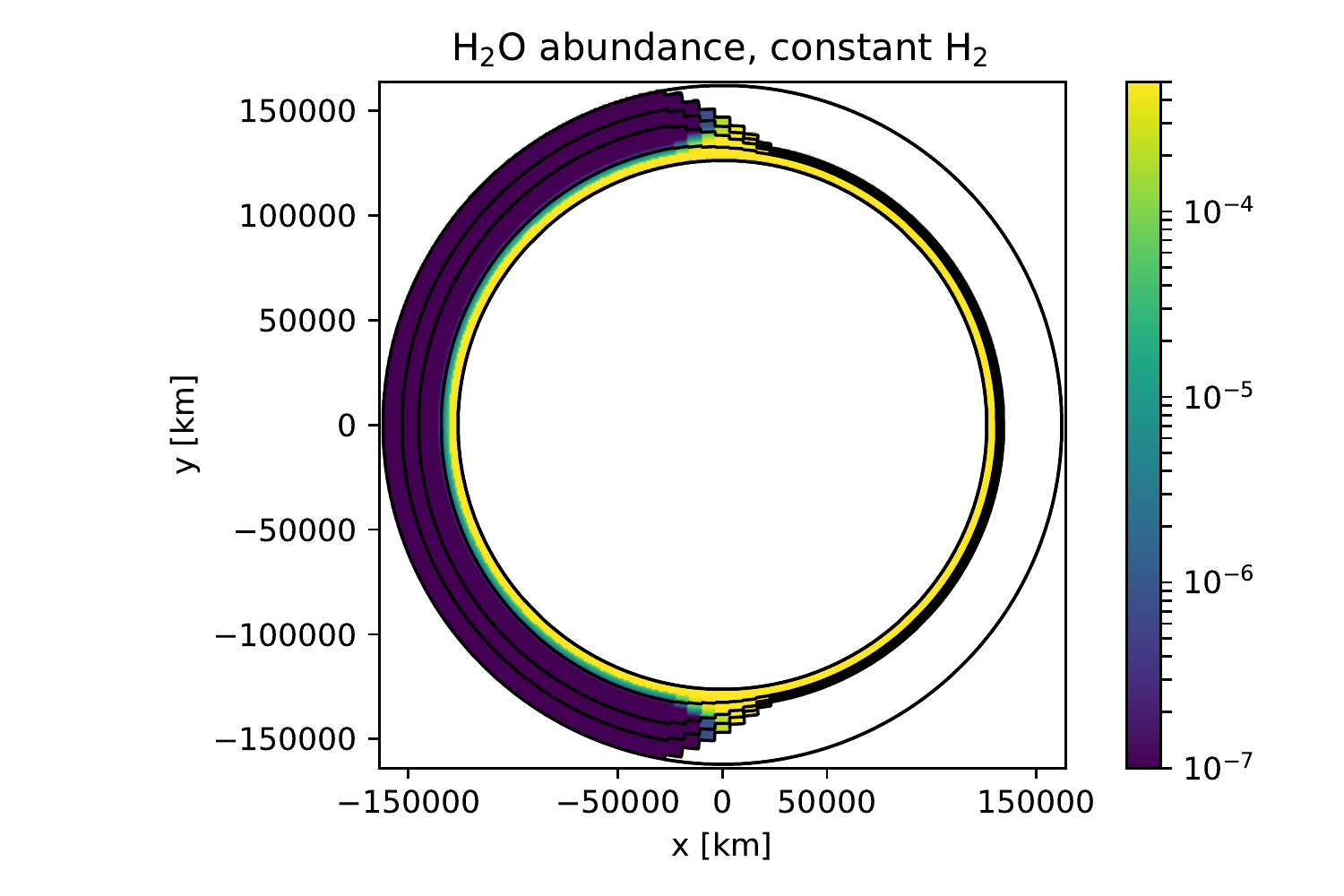}
\includegraphics[scale=\sizefigTp,trim = 1.5cm 0cm 0.cm 0.cm, clip]{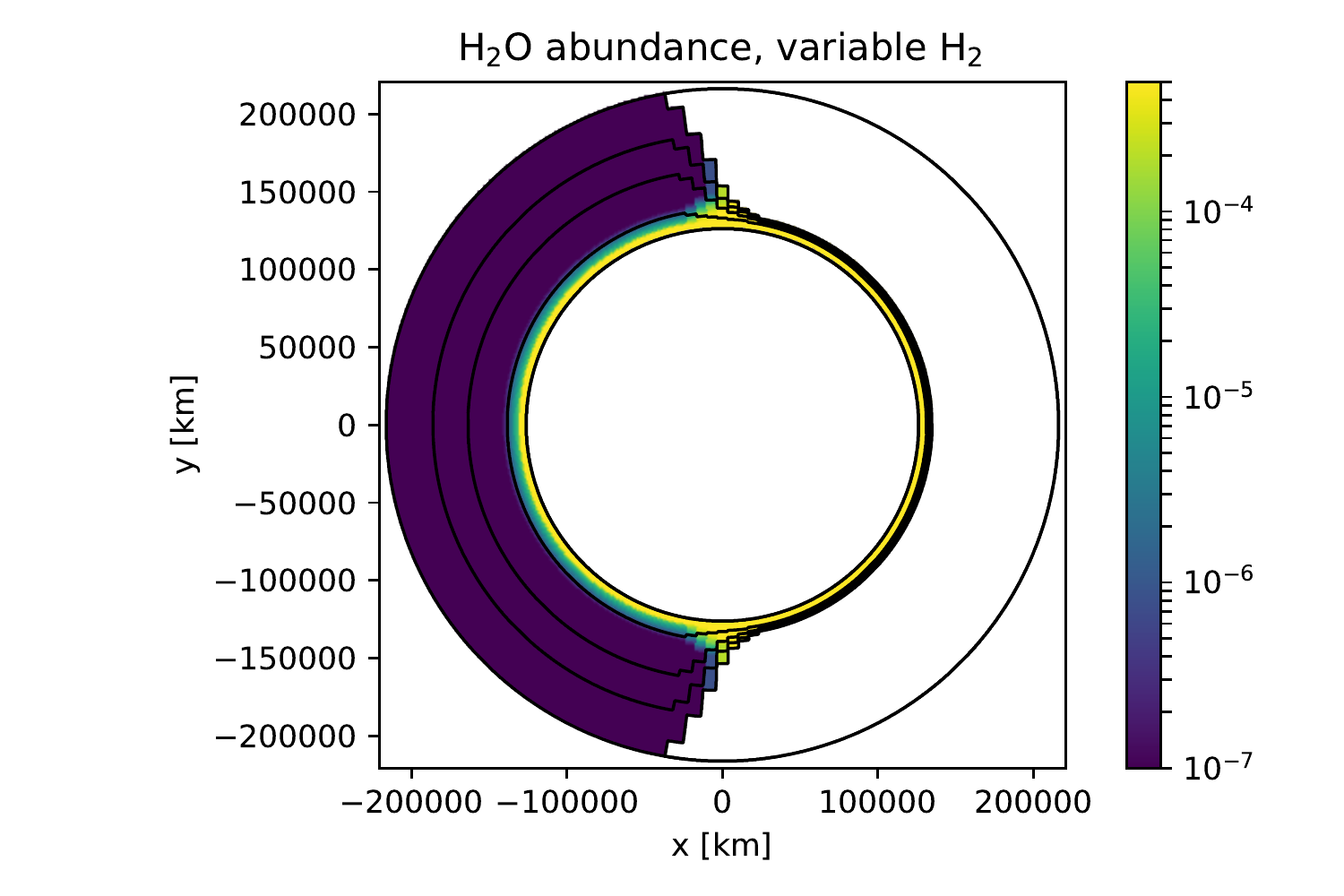}
\caption{Atmospheric structure of the symmetric, idealized case \Tp, assuming an absence of \hh\ dissociation. We show the temperature map (top) and the water abundance map (bottom). The temperature gradient goes from $\mathrm{500\,K}$ to $\mathrm{3500\,K}$. The transmission angle is 10° around the terminator line and have a $\mathrm{2500\,K}$ ring at P=$\mathrm{0.13\,bar}$ from the surface pressure. From center outward, the 5 solid lines are, respectively, the $10^6$, $10^3$, 1, $10^{-2}$, and $10^{-4}$ Pa pressure levels. The colormap for water abundance maps goes from $5.10^{-4}$ to $10^{-7}$. Note that the radius of the planet and the atmosphere are shown to scale. The temperature and abundance maps of the \Tm case are almost the same with only a smaller scale height of the atmosphere.}
\label{fig: Atmospheric structure symcase}%
\end{figure*}

\fig{fig: Atmospheric structure Tm} and \fig{fig: Atmospheric structure symcase} show the temperature structure of the atmosphere of the cases studied. These maps are plotted to scale. Note that we consider a symmetric atmospheric structure, so the equatorial cut and the polar cut are exactly the same. As we can see, the day side of those atmospheres are strongly inflated in comparison to the night side which will create the horizontal biases that we want to characterize.

The \Tp and \Tm cases have been chosen to encompass the real day-night gradient present in \wasp (see below) while removing effects due to differences along the limb (for example, east vs. west limb or equator vs. poles). This should allow us to identify the specific effect of these differences by comparing the retrieval results with our idealized cases and with the actual GCM outputs.

The \Tp case is close to the GCM case. \citet{BC18} and \citet{TK19} showed that the energy redistribution effect of hydrogen recombination cools the day side, heats the night side and reduces the gradient across the limb. The \Tm case thus uses a reduced gradient to account for this effect and present a very conservative estimate of the 3D effect.

\begin{figure}
\centering
\includegraphics[scale=0.37]{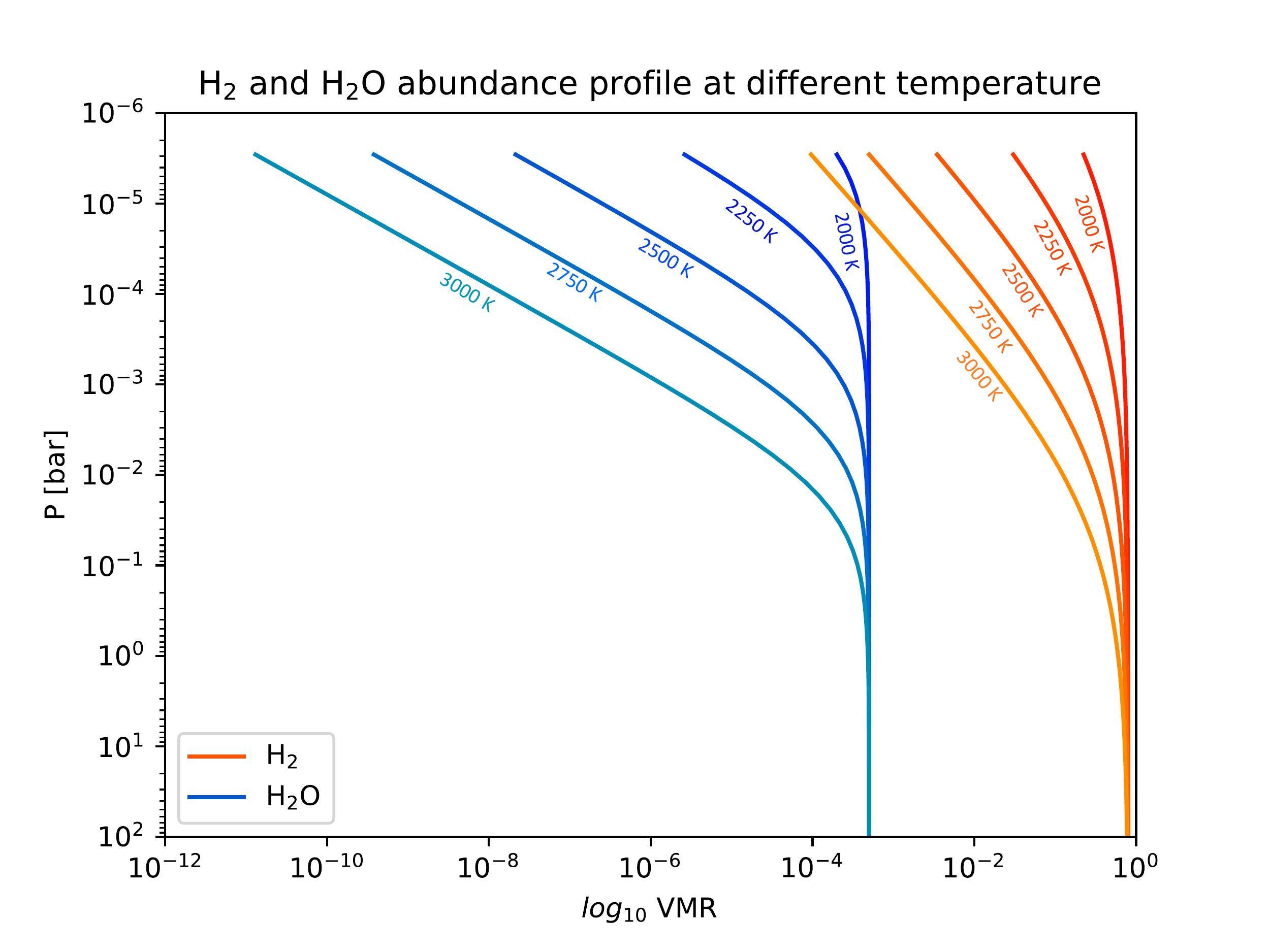}
\caption{H$_2$O and H$_2$ abundance profiles plotted for a range of temperature between 2000 K and 3000 K. At a given pressure, H$_2$O and H$_2$ dissociation depends strongly on the temperature. The higher the temperature, the lower the abundance in the atmosphere. At a fixed temperature, the variation of the dissociation rate depend as a function of P and $\mathrm{P}^2$, respectively, for H$_2$ and H$_2$O.}
\label{fig: Water_H2_abundances}
\end{figure}

\subsubsection{Composition effects due to the dissociation of some molecules}

In UHJs, molecular abundances are not the same everywhere, since the day-side temperature is high enough to allow thermal dissociation of many species \citep{Lodders2002,Visscher2006,Visscher2010,Marley2017}. In this work, we take into account the dissociation of two molecular species, first only \hho\ and later also \hh, using the equations provided by \citet{Parmentier2018} as shown in \fig{fig: Water_H2_abundances}. At a given temperature, the lower the pressure, the stronger the dissociation and at a given pressure, higher the temperature higher the dissociation. In this study, we assume that dissociated \hh\ predominantly forms atomic hydrogen. We know that hydrogen anion can absorb part of the electromagnetic radiation between 1$\mu$m and 10$\mu$m \citep{Lenzuni1991,Parmentier2018}. However, we did not include in our computations the absorption contribution from ionized hydrogen. The main reason is that we want to be consistent with TauREx, which does not include this opacity source. Furthermore, the dissociation of \co$\,$ cannot occur at the temperature regimes in our simulations, i.e. 500K to 3500K. In order to dissociate, \co$\,$ requires higher temperatures, due to its stronger triple bond \citep{Lodders2002}. For this reason, we assumed in all the simulations, a constant \co$\,$ abundance everywhere in the atmosphere.
In the first set of simulations, we considered the \Tp and \Tm models considering water dissociation alone (\hh\ is held constant). As shown in the abundance map of \hho\ in \figs{fig: Atmospheric structure Tm}{fig: Atmospheric structure symcase}, there is no more water in the day-side of the planet at pressure lower than $\Piso$. The dissociation stops around the terminator and, in the night-side, the water distribution comes back at a constant value.

In the second set of simulations we took into account both \hh\ and \hho\ dissociation. As \hh\ does not have significant absorption bands in the 0.6 - 10$\mu$m wavelength range, the main effects of its dissociation is to decrease the molecular weight, causing an increase of the atmospheric scale height. It also affects the thermochemical properties of the atmosphere because atomic hydrogen recombines on the night side. This redistribution of energy will heat the night side and cool the day side \citep{BC18,TK19}. An other important effect is that without \hh\ in the atmosphere, collision induced effects from both, \hh-\hh$\,$ and \he-\hh$\,$ collisions, are less significant. As shown in \figs{fig: Atmospheric structure Tm}{fig: Atmospheric structure symcase}, \hh$\,$ dissociation increases the atmospheric scale height of the day side by a factor of 1.5 and 2 in the \Tm and \Tp simulations, respectively. We note that, the temperature is not high enough to induce \hh\ dissociation in the night side. This characteristics, creates an even stronger difference between the night side and the day side when \hh\ dissociates. Consequently, \co\ is the dominant radiatively active species in the day-side because it does not dissociate at these temperatures.

\subsection{3D models of \wasp atmosphere using SPARC/MITgcm global circulation model}
\label{ssec:GCM}

In the last configuration, we add both the vertical and azimuthal (along the limb) effects. Indeed, we model \wasp using the Global Circulation Model (GCM) SPARC/MITgcm \citet{Parmentier2018} described in Sect \ref{sec:GCM}. The chemistry is kept unchanged. As in \sect{ssec: 3D Symmetric}, we studied four different cases considering all the possible combinations where water and \hh\, dissociate.
\figs{fig: Atmospheric structure}{fig: Atmospheric structure H2 diss} show the temperature and water abundance maps for the \gcm cases with and without \hh\ dissociation. We see that the thermal structure of the atmosphere is more complex than the one in the symmetric case, even if the salient features remain the same. The most important modification concerns the terminator which is now asymmetric because of the winds created by the strong temperature dichotomy between the day and night side \citep{Showman2015}. Those equatorial jets warm the east side of the atmosphere and cools the west side around the equator, which implies less dissociation of water on the west because of the cooler temperature. Nevertheless, the abundance maps show that there is still water dissociated at the terminator which could allow the light to probe deeper into the atmosphere on the night side of the planet. Note also that when \hh\ dissociation is taken into account, the global behavior of the 3D structure remains the same, but since the day side of the atmosphere has a bigger scale height, it is more inflated.

An important caveat here is that while we model \hh\ dissociation to compute the transit spectrum, the dissociation is not taken into account in the GCM itself. This is expected to reduce the day night contrast thus the winds fields \citep{BC18,TK19}. So this case is not completely self consistent. 


\begin{figure*}
    \centering
    \begin{subfigure}[b]{0.47\textwidth}
        \includegraphics[scale=0.55]{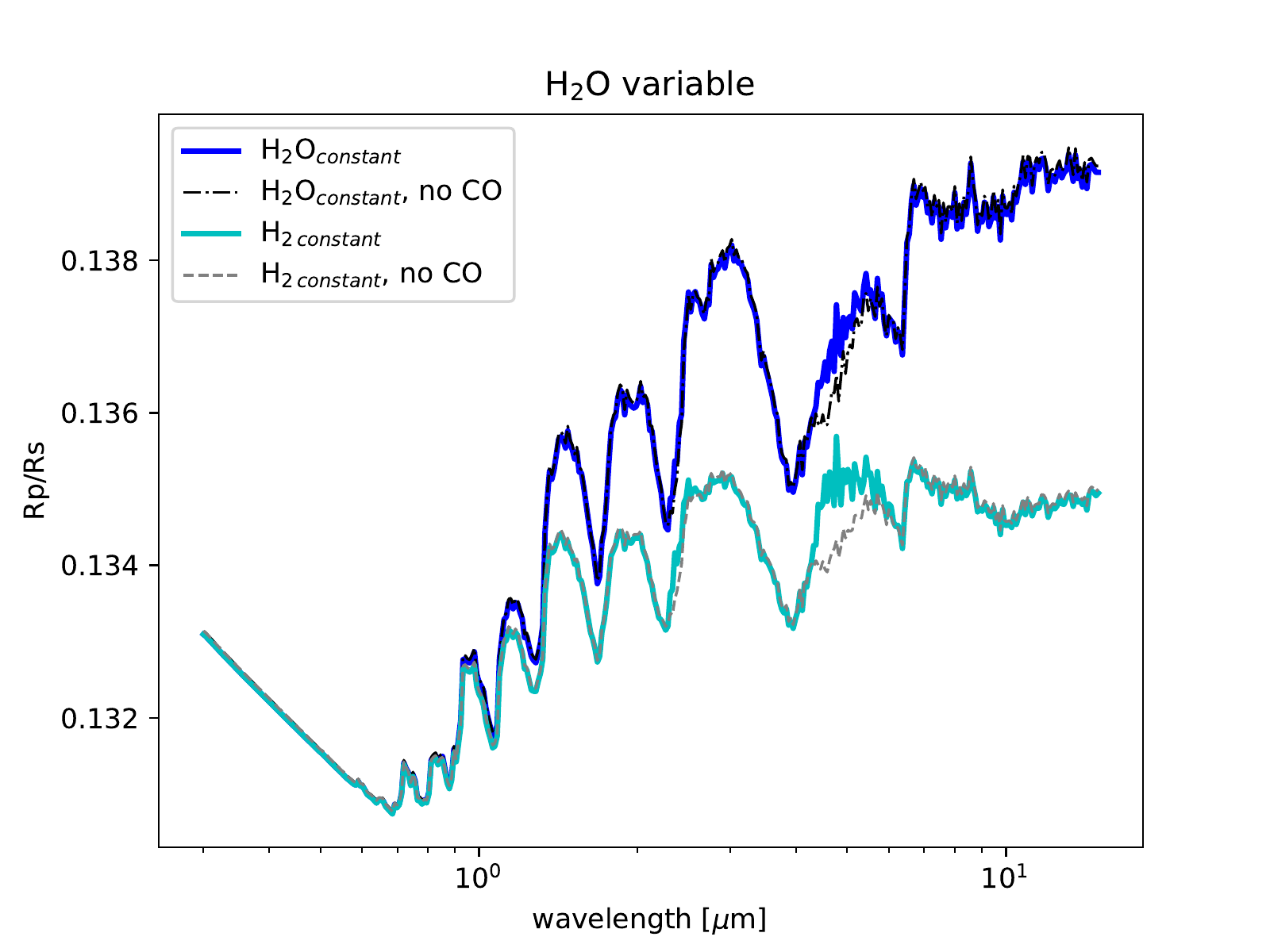}
        \caption{}
        \label{fig:H2cst}
    \end{subfigure}
    \begin{subfigure}[b]{0.47\textwidth}
        \includegraphics[scale=0.55]{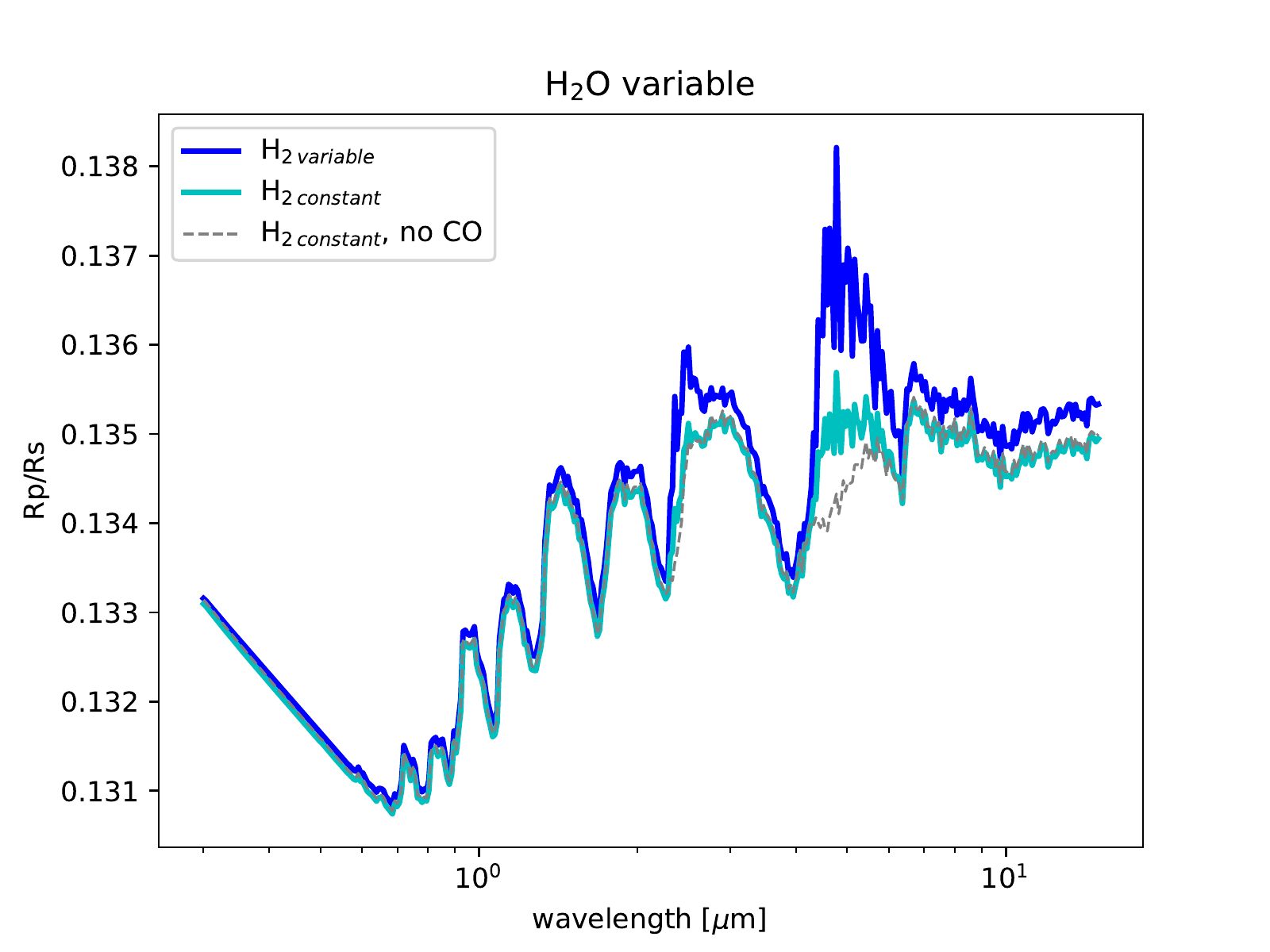}
        \caption{}
        \label{fig:H2var}
    \end{subfigure}
    \caption{(Left): Transmission spectra of \wasp at resolution of R = 100 for \gcm simulations assuming a constant \hh abundance in the whole atmosphere. When water dissociation is taken into account (light blue line), the water features become shallower compared to when we assume no water dissociation (blue line). (Right): Transmission spectra of \wasp at resolution of R = 100 for \gcm simulations taking into account \hho dissociation in the atmosphere. When \hh\ dissociation is considered (blue line), the \co\ features appears more clearly compared to when we neglect \hh\ dissociation (light blue line). Black line and grey line correspond to the transmission spectra for an atmosphere without \co\ for the water constant and dissociated case respectively to highlights the features of \co\ in the other curves.}
    \label{fig: Spectra_H2cst_gcm}
\end{figure*}

\begin{figure*}
\centering
\includegraphics[scale=\sizefigHfix,trim = 0cm 0cm 0.cm 0.cm, clip]{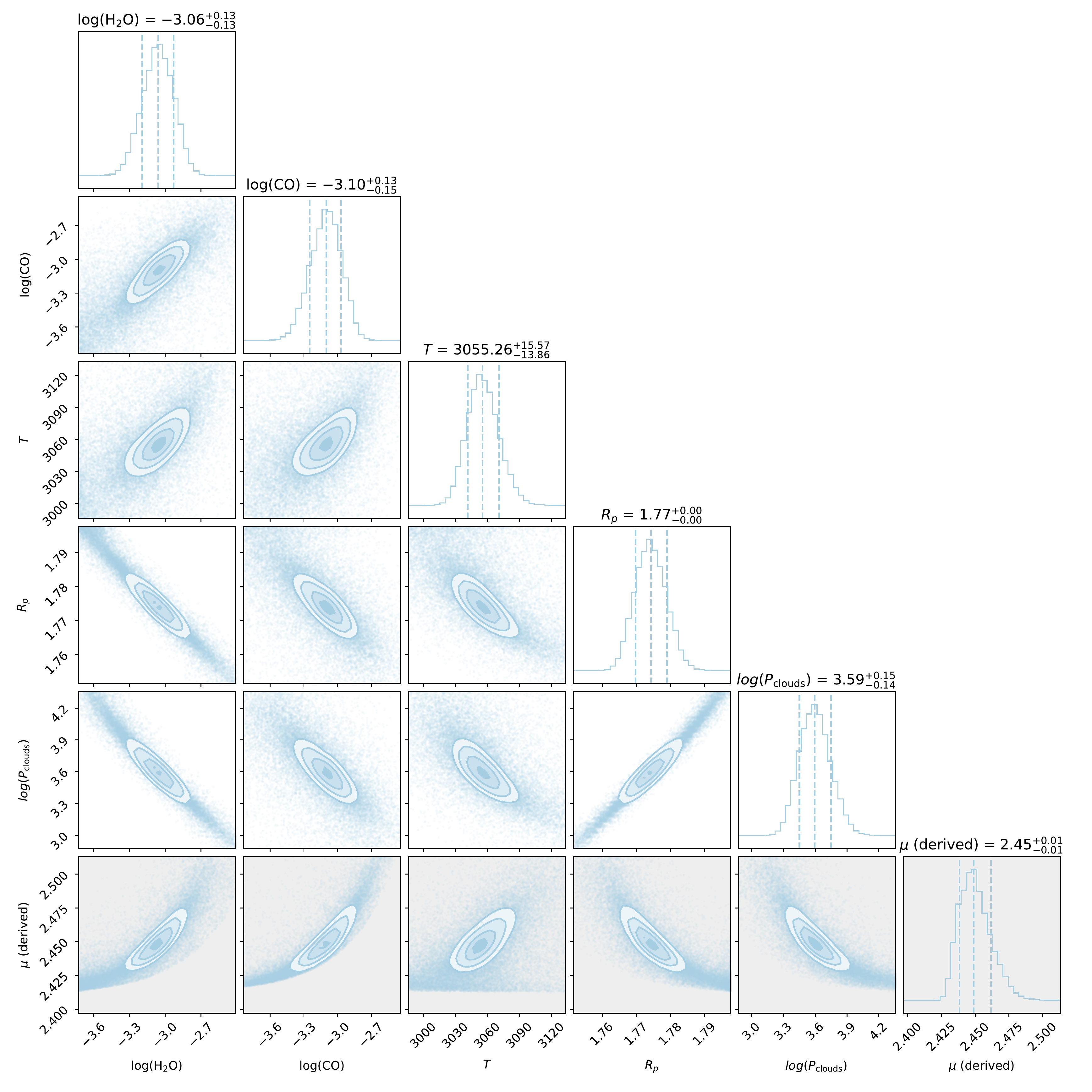}
\includegraphics[scale=\sizefigHfix,trim = 0cm 0cm 0.cm 0.cm, clip]{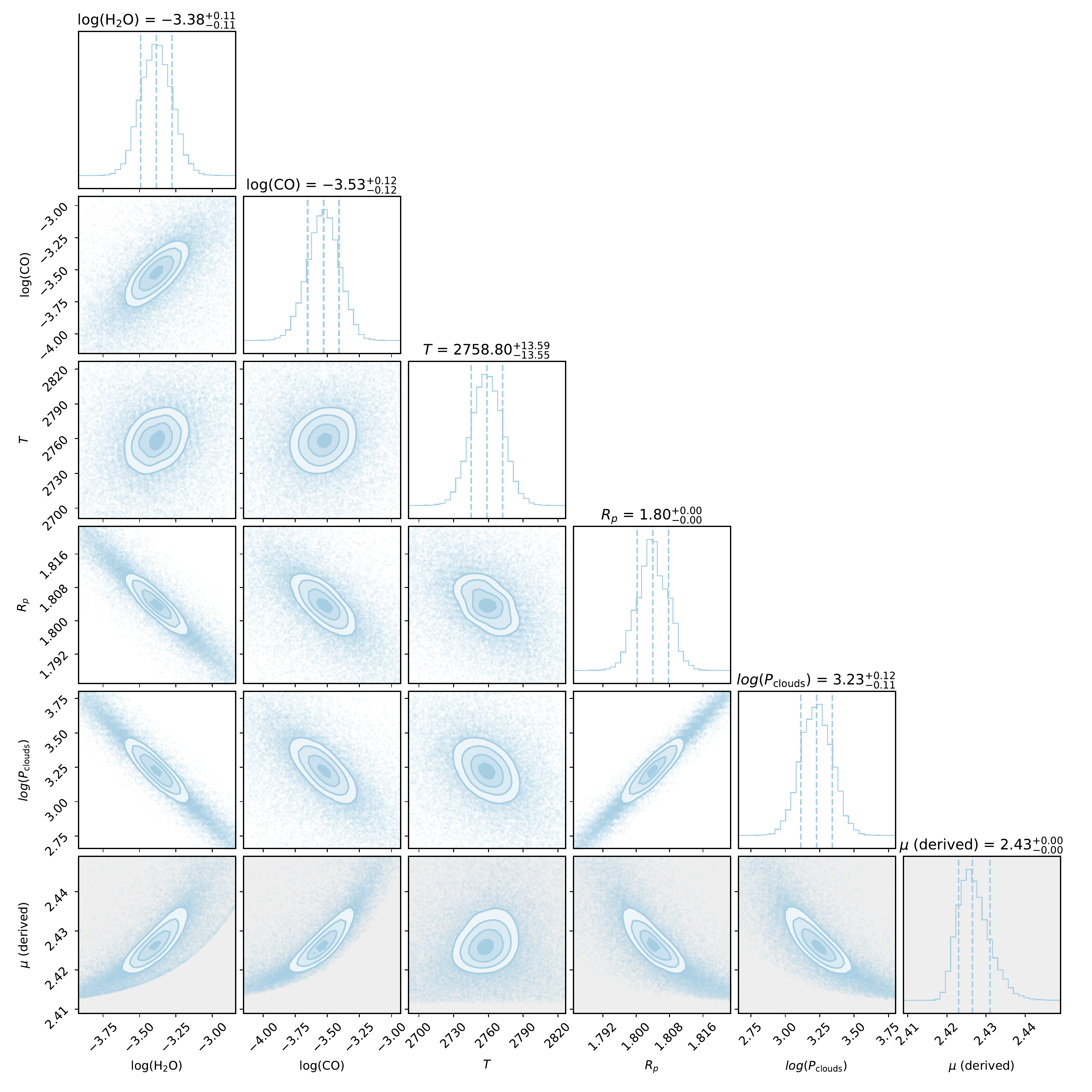}\\
\includegraphics[scale=\sizefigHfix,trim = 0cm 0cm 0.cm 0.cm, 
clip]{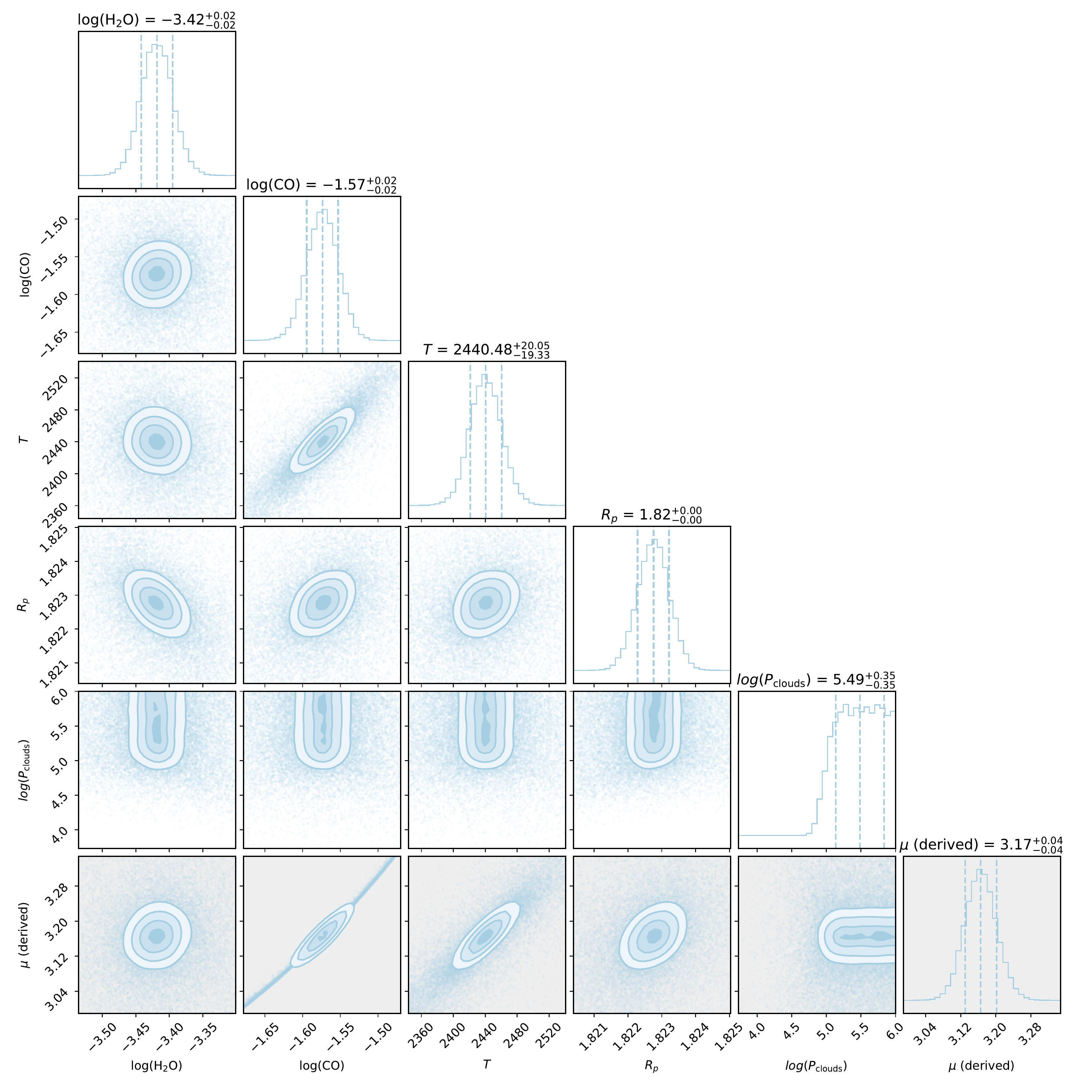}
\includegraphics[scale=\sizefigHfix,trim = 0cm 0cm 0.cm 0.cm, clip]{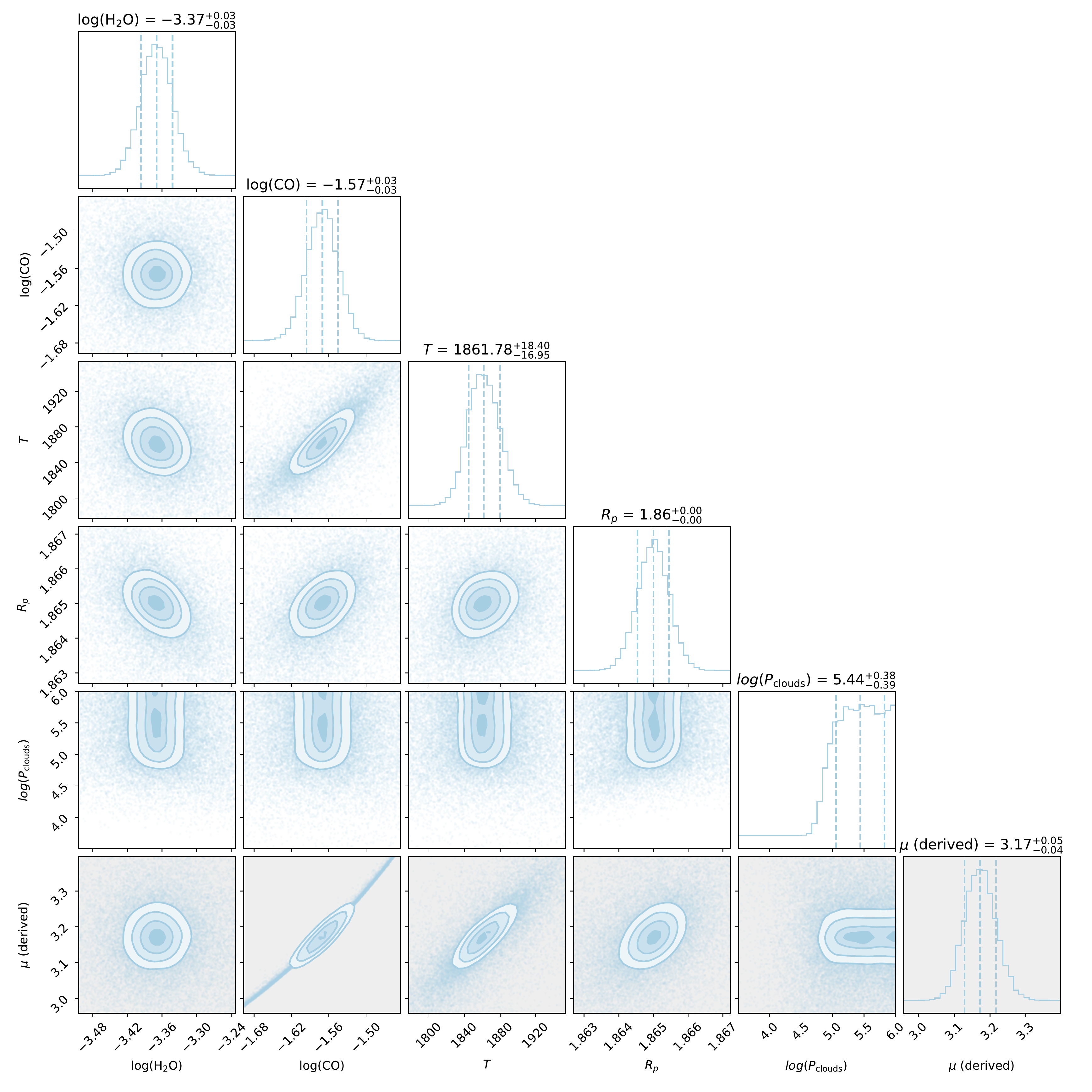}
\caption{Nest posteriors retrieval from \taurex assuming a constant \hh\ abundance in the atmosphere. There are the \Tp (left) and \gcm cases (right) which assume, respectively, no water dissociation (Top) and water dissociation (Bottom). We retrieved five free parameters which are \hho, \co\ abundances in log10(VMR), clouds pressure in bar, the temperature in Kelvin and the planetary radius in Jupiter's radius. The molecular weight is derived from those parameters.}
\label{fig: H2fix_nest_posteriors}
\end{figure*}


\begin{figure*}
\centering
\includegraphics[scale=\sizefig,trim = 0cm 0cm 0cm 0cm, clip]{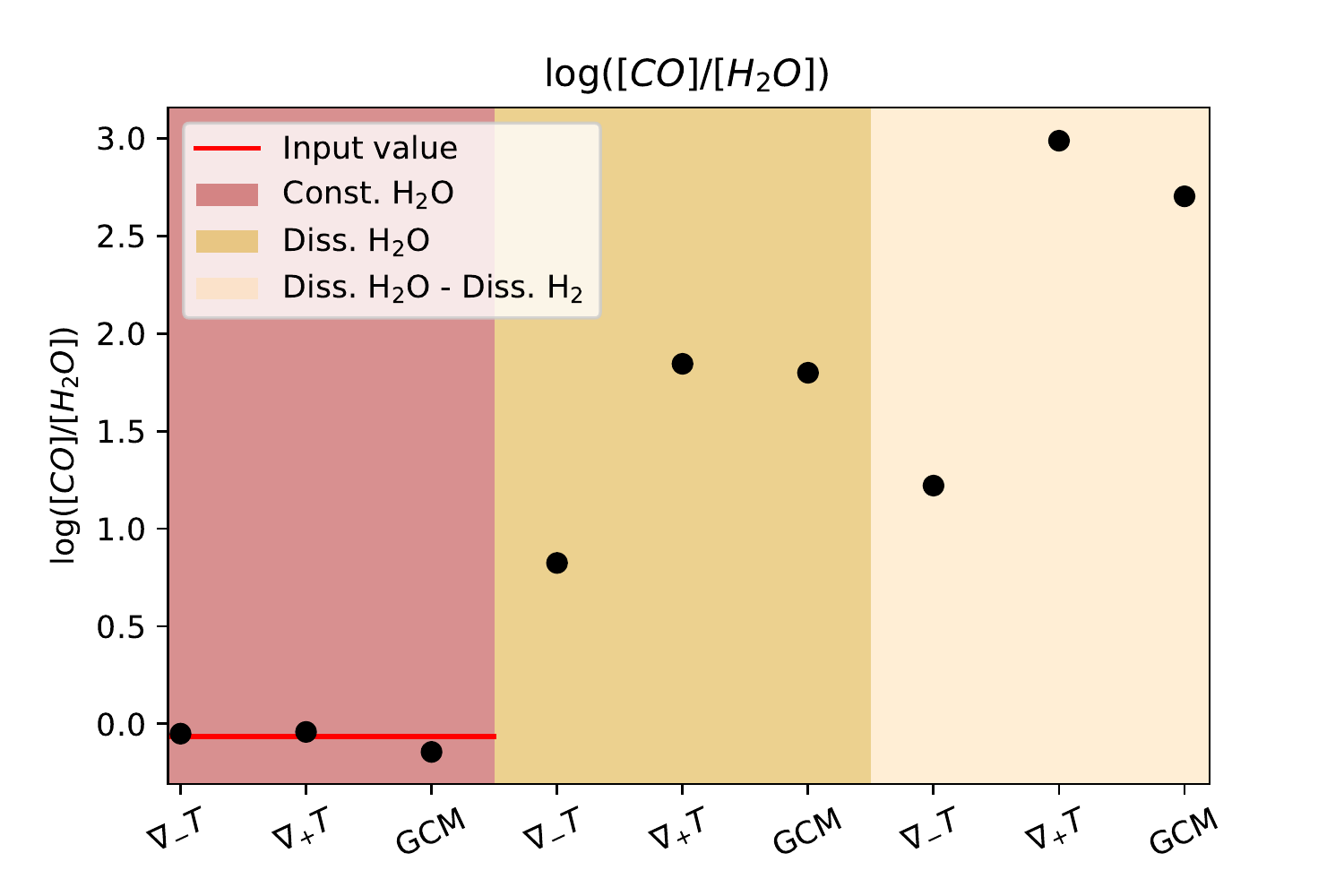}
\includegraphics[scale=\sizefig,trim = 0cm 0cm 0cm 0cm, clip]{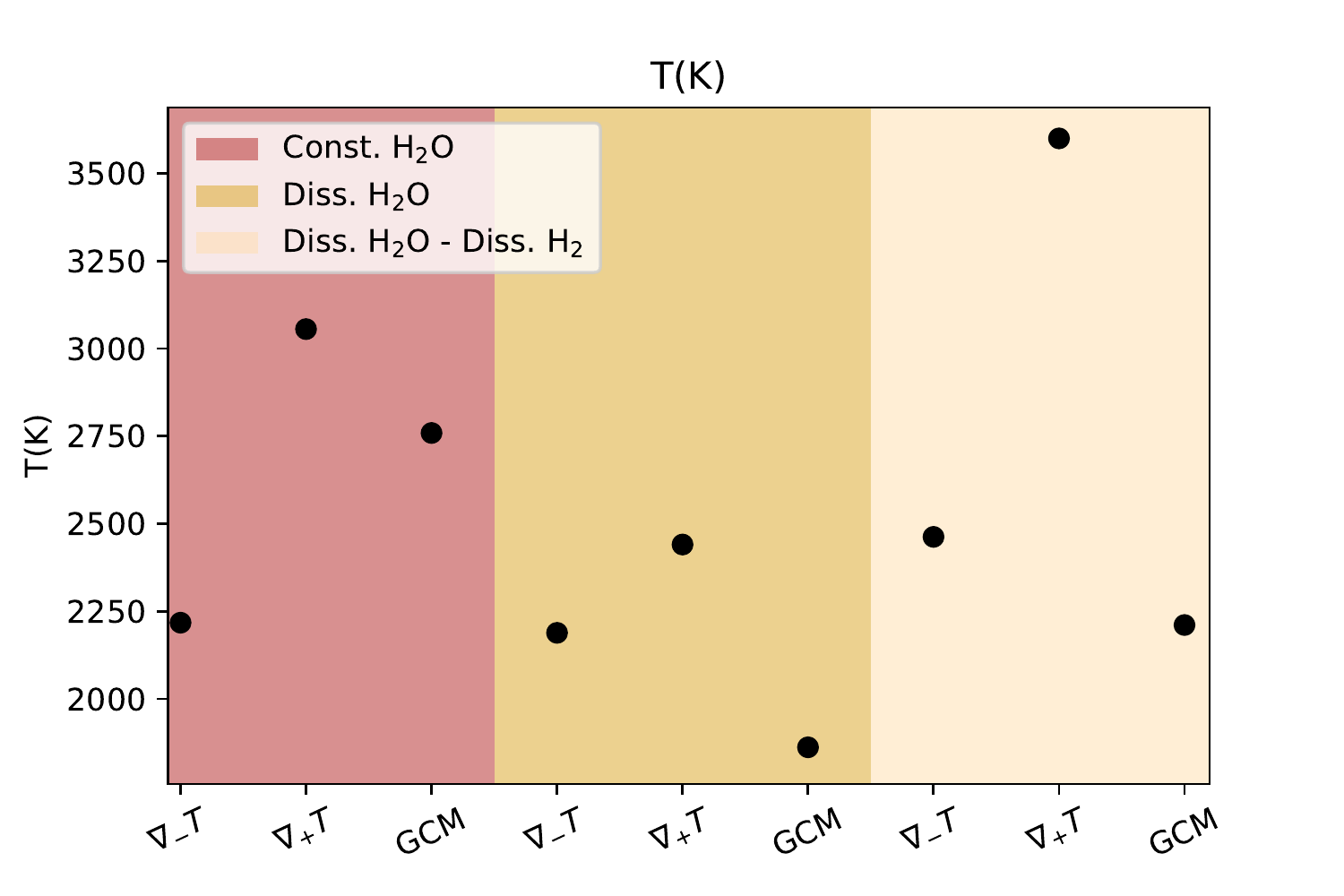}\\
\includegraphics[scale=\sizefig,trim = 0cm 0cm 0cm 0cm, clip]{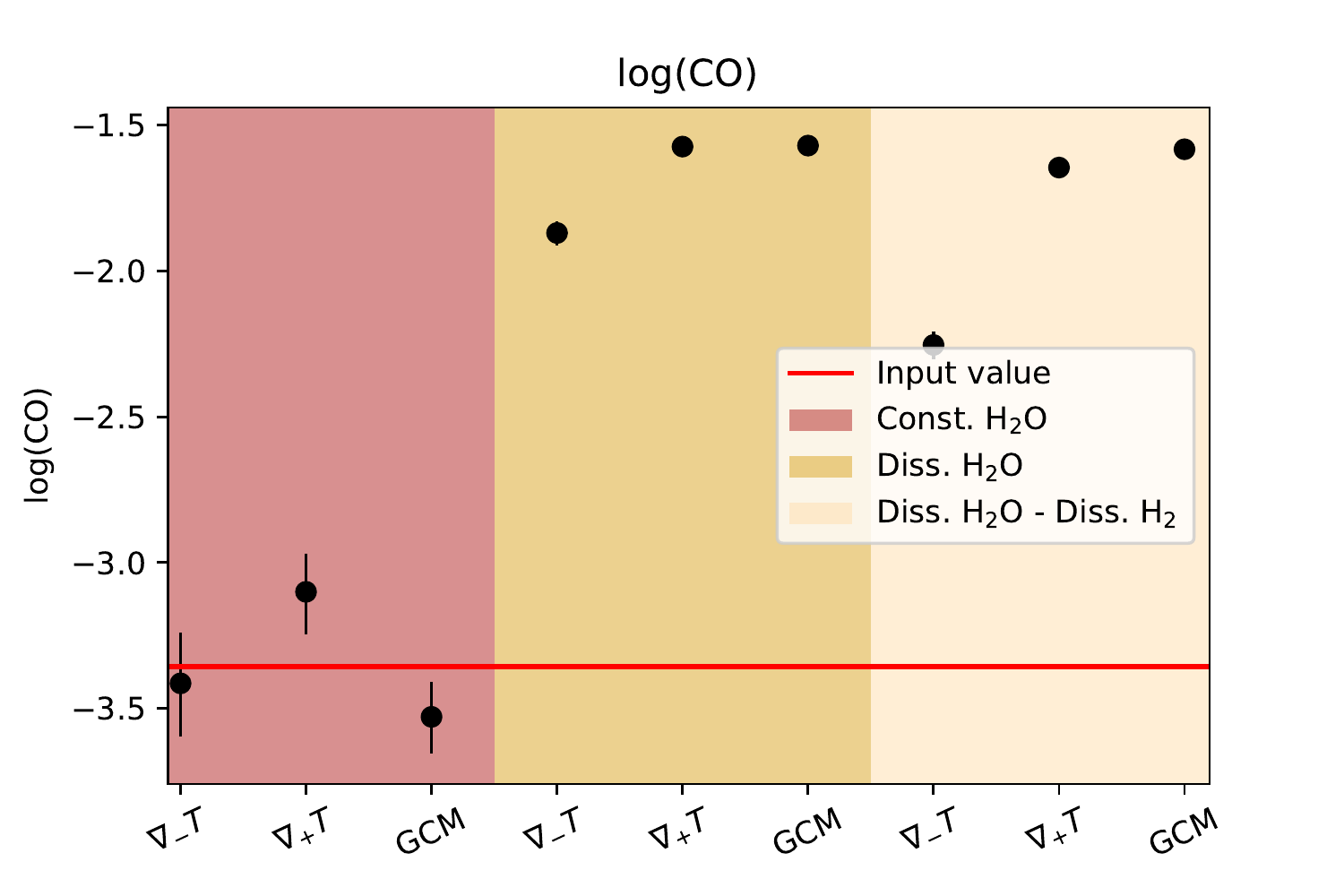}
\includegraphics[scale=\sizefig,trim = 0cm 0cm 0cm 0cm, clip]{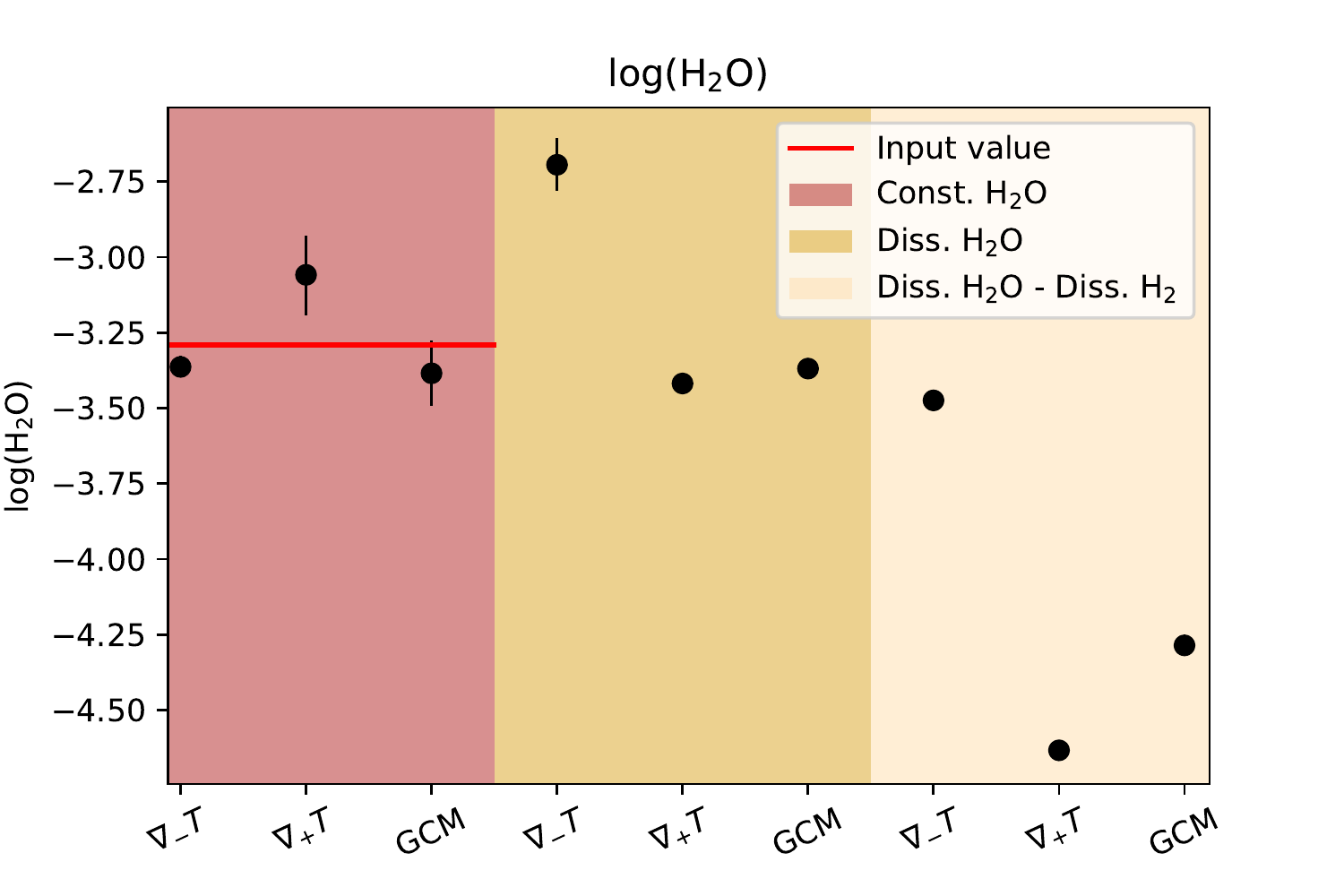}\\
\includegraphics[scale=\sizefig,trim = 0cm 0cm 0cm 0cm, clip]{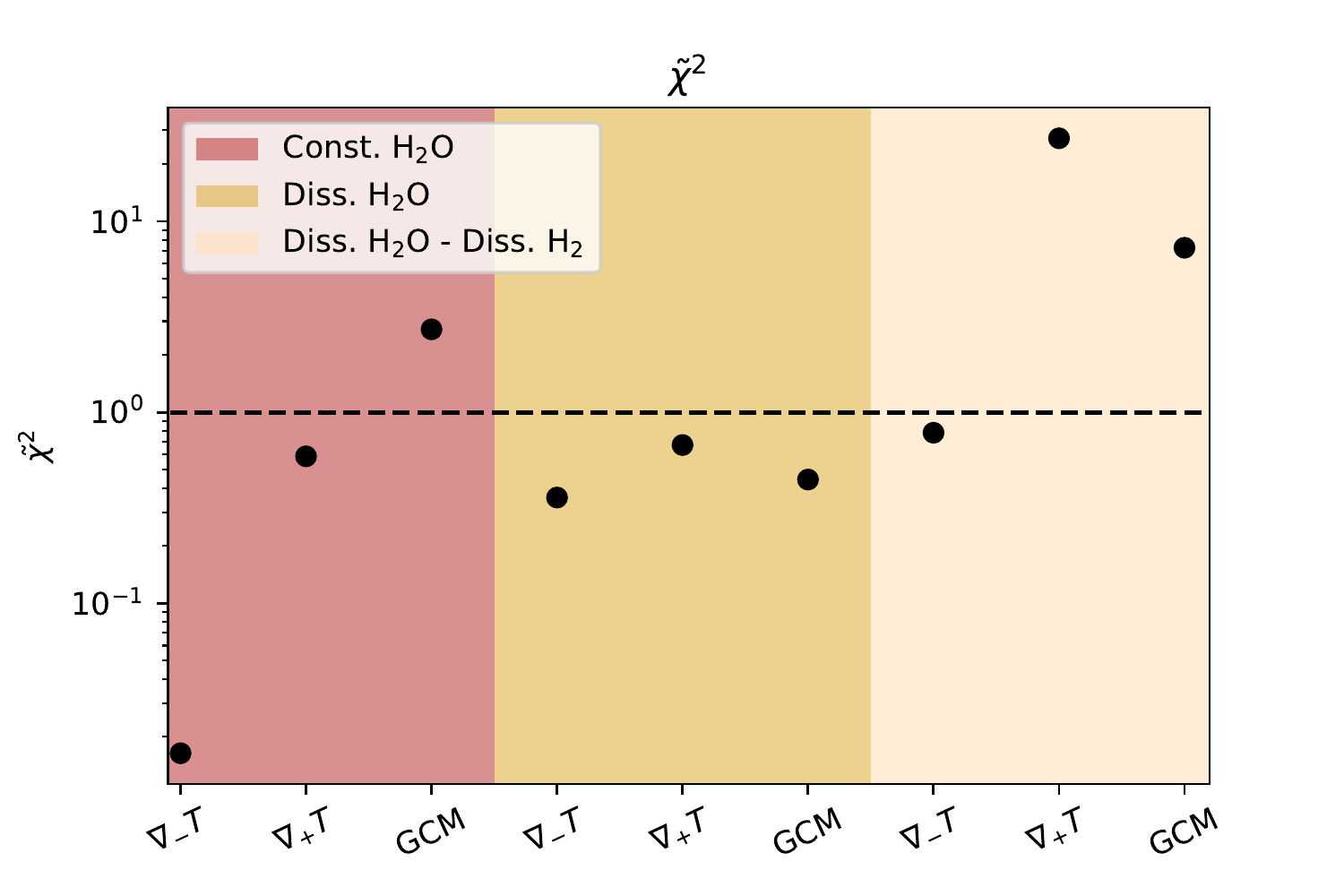}
\includegraphics[scale=\sizefig,trim = 0cm 0cm 0cm 0cm, clip]{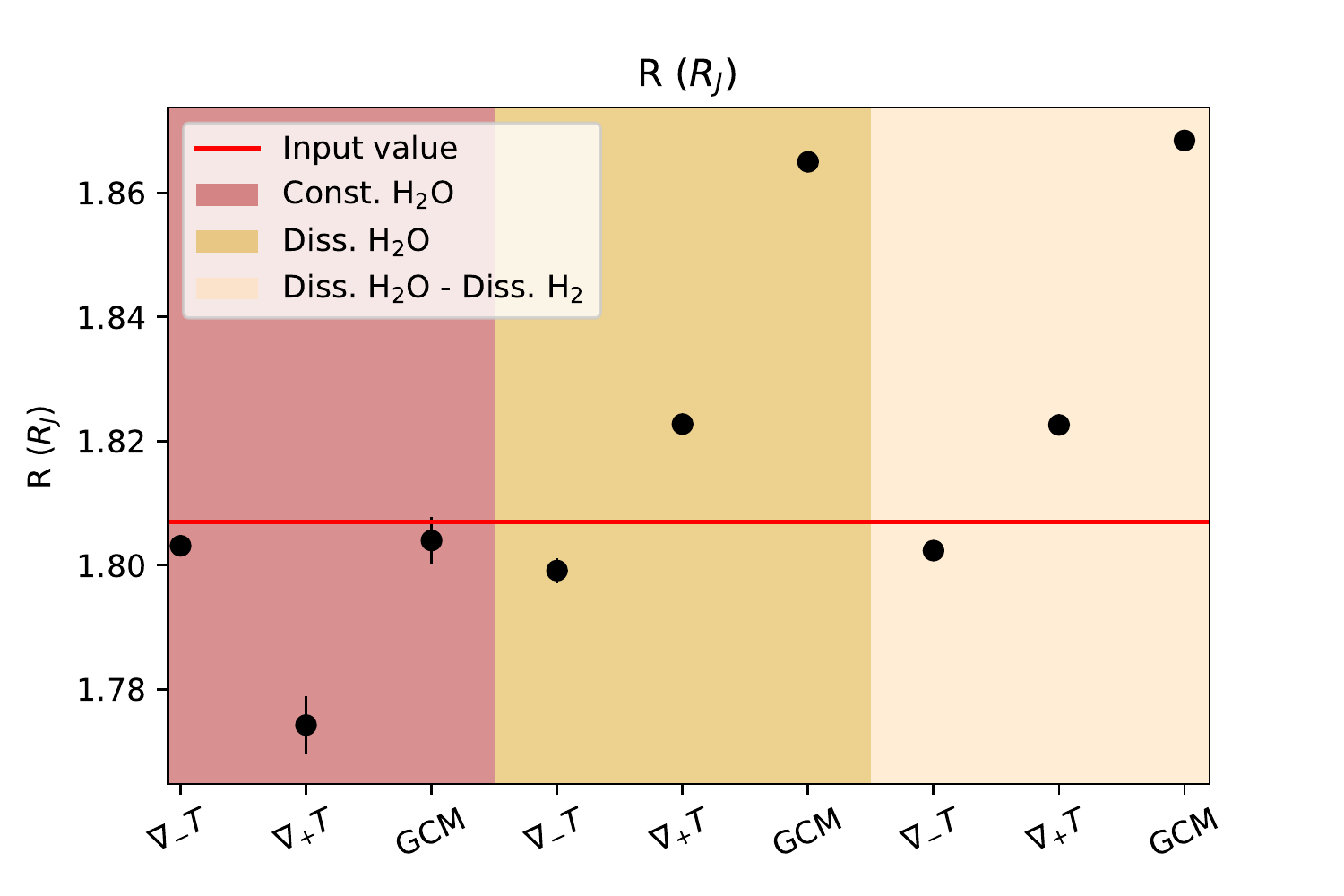}
\caption{Summary of all the retrieval results for \Tm, \Tp and \gcm  simulation considering every cases. We show the \COratio ratio (top left), the temperature (top right), the log abundances of \co\ (middle left) and \hho\ (middle right), the \redchi\ (bottom left) and the planetary radius (bottom right). Those retrievals have been calculated with a shot noise assuming a floor noise of 30ppms through the whole spectral domain. The red line represents the input value from our simulations and the black dot line shows where the \redchi=1.}
\label{fig: comparison plot summarize}
\end{figure*}

\begin{figure}
\centering
\includegraphics[scale=0.64,trim = 0cm 4cm 1.5cm 0cm, clip]{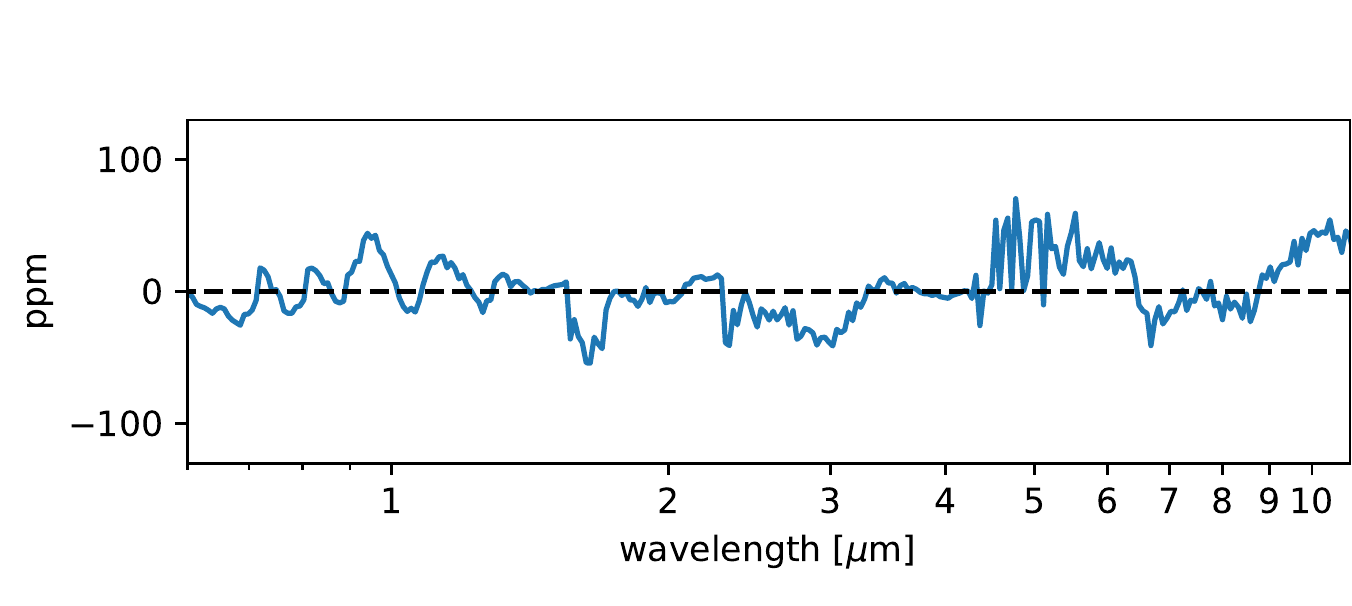}
\\
\includegraphics[scale=0.49,trim = 0cm 0cm 2cm 1.5cm, clip]{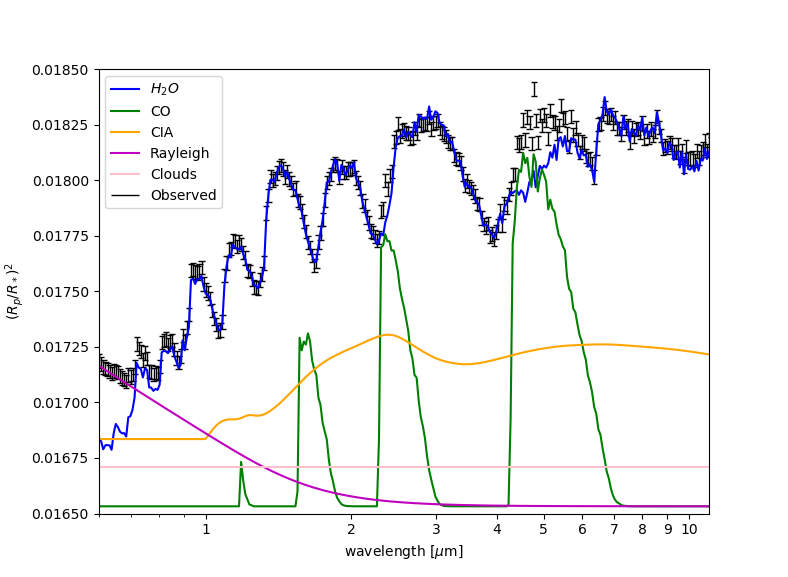}
\caption{Particular solutions found by TauREx in the \gcm case assuming water dissociation and \hh\ constant. In the bottom: we show the contribution plots of each component, i.e. H$_2$O, CO and Collision Induced Absorption (CIA) for the solution found by TauREx. In the top: residuals (in ppm) between the input spectrum and the best-fit spectrum.}
\label{fig: contribution-gcm}
\end{figure}

\section{Results}
\label{sec:Results}


In this section, we describe the results of the transmission spectra generated by \pyt and the spectral retrieval using our three set of simulations, each one being performed on our three temperature structures (\Tm, \Tp, and GCM):
\begin{enumerate}
\item H$_2$ and H$_2$O constant in the whole atmosphere;
\item Thermal dissociation of H$_2$O with H$_2$ held constant;
\item Thermal dissociation of both H$_2$O and H$_2$.
\end{enumerate}

\subsection{The 3D transmission spectrum of ultra hot Jupiters}
\label{3D trans spectra}

The transmission spectra generated by \pyt are shown in Fig. \ref{fig: Spectra_H2cst_gcm}. Here, we tested two cases with \hh\ constant considering or not water dissociation in the atmosphere (Fig. \ref{fig: Spectra_H2cst_gcm}a), and a third one with both \hho\ and \hh\ dissociation (Fig. \ref{fig: Spectra_H2cst_gcm}b). 
When \hho\ is constant, the water features are more evident and the \co's features are barely visible as it is highlighted comparing the transmission spectra with and without \co\ in the atmosphere (respectively the blue and the dotted black lines). Since there is \co\ and \hho\ everywhere, then the transmitted light probes the same region for both molecules, which are at a high altitude in the day side of the atmosphere. In this case then the spectral features are similar to the ones of the 1D case.

When \hho\ dissociates (light blue curve in \fig{fig: Spectra_H2cst_gcm}a), the water features in the spectrum become smaller because it almost disappears from the day side of the atmosphere. Then, the light from the star can probe much deeper regions of the atmosphere in the wavelength range where the \co\ does not absorb and reaches part of the night side. The \co, which is not dissociated on the day side - where the scale height is large due to the high temperature - appears more evidently. So, as the light goes first through the day side and then through the night side, the transmission spectrum carries information of both sides of the atmosphere depending on the wavelength we look at. The behavior of the spectrum can also be understood looking at \figs{fig: Atmospheric structure}{fig: Atmospheric structure symcase} where the maps of the atmosphere show the lack of water in the day side except for a deep part of the atmosphere ($\simeq\,$10$\,$mbar). Since \co\ does not dissociate, it absorbs mainly in the day-side of the atmosphere, where the high temperature leads to larger absorption features.

When both \hho\ and \hh\ dissociate, the water features are similar in both presence and absence of \hh\ dissociation as shown in Fig \ref{fig: Water_H2_abundances} and Fig \ref{fig: Spectra_H2cst_gcm}b. Indeed, the main effect of the dissociation of \hh\ in the atmosphere is the increasing of the scale height. As the dissociation of \hh\ mainly occurs in the day side of the atmosphere as well as the dissociation of \hho\ (see Fig \ref{fig: Atmospheric structure} and \ref{fig: Atmospheric structure H2 diss}) the region probed is the same in both cases.
However, the \co\ features appears even more clearly. As \co\ abundance remains constant everywhere in the atmosphere, its features are more significant due to the enlarger of the scale height caused by the dissociation of \hh\ in the day side.

Note that \fig{fig: Spectra_H2cst_gcm} represents the transmission spectra for the case with the \gcm input, but the whole behaviours explained above remain similar to the symmetric \Tp and \Tm cases.

\subsection{Retrieval results}

\subsubsection{\hh\ and \hho\ constant in the atmosphere}
\label{h2oconst-h2const}

The posterior distributions of \Tp and \gcm simulations are shown in Fig \ref{fig: H2fix_nest_posteriors}. The retrieval code \taurex converges, in every simulation, to a non-degenerate solution. \fig{fig: comparison plot summarize} summarizes all the solutions found by \taurex for each simulation.
As shown in the red part of the various panels of this figure, \taurex finds non-degenerate solution within 2$\sigma$ from the input H$_2$O and CO abundance for the \Tp, \Tm and \gcm simulations. Those chemically homogeneous simulations are similar to what \citet{Caldas2019} have done, that's why we expected to find consistent results with their work.
To better interpret those results, we plotted the \COratio ratio (Fig~\ref{fig: comparison plot summarize}). This ratio allows us to understand if the dissociation of water affects the spectrum. The \COratio ratio is constant when water does not dissociate, which it is found here by \taurex retrieval.
The temperatures retrieved suggest that the light is probing the day side of the atmosphere. Indeed, as shown in Fig \ref{fig: Atmospheric structure symcase} and Fig \ref{fig: Atmospheric structure}, the atmosphere presents an inflated day side. The temperature retrieved in the \Tp case is higher than the one retrieved in the \Tm case which is expected since the temperature of the day side are higher in the \Tp case. Using as input the spectrum computed with GCM, \taurex finds a slightly lower temperature than the one found in the two symmetric simulations. Considering that the GCM simulations has been made assuming a non-isothermal temperature-pressure profile, the temperature structure of GCM is indeed more complex than the one in the other two simulations. These results are all consistent with \citet{Caldas2019}. Note that the calculated $\chi^2$ is much higher for the \gcm case compared to the symmetric cases \Tp and \Tm.

\subsubsection{Thermal dissociation of \hho\ with \hh\ constant}
\label{h2ovar-h2const}


When water dissociation is taken into account, TauREx returns significantly different solutions compared to the homogeneous simulations. The results are shown in the yellow parts of \fig{fig: comparison plot summarize}.

First, all the retrieved temperatures are lower than the ones found in the constant composition simulations. Because the main constraint that allows \taurex to retrieve the temperature is the amplitude of the spectral bands - the higher the temperature the higher the amplitude - the temperatures retrieved are lower because the water features are coming from a colder region of the atmosphere. In the \gcm case, the temperature retrieved is lower than the temperatures of the limb, which suggest that we probe the night side of the atmosphere. 

Since \co\ does not dissociate, it absorbs mainly in the day-side of the atmosphere, where the high temperature leads to larger absorption features as explained in Sect \ref{3D trans spectra}.
Looking at the yellow part of Fig \ref{fig: comparison plot summarize}, there are non-degenerate solution for \taurex to fit the \co\ features, considering that it also finds a low temperature because of the presence of water in the cold night-side of the planet.
This solution suggests a high \co\ abundance in the atmosphere and converges around the correct water abundance within 3$\,\sigma$. This water abundance corresponds indeed to the abundance in the night side of the atmosphere where water is not dissociated (see Figs \ref{fig: Atmospheric structure}, \ref{fig: Atmospheric structure Tm} and \ref{fig: Atmospheric structure symcase}).

As a consequence, the \COratio is around 100 times higher than expected. We note that for the conservative case \Tm\, the \COratio is less biased, around 10 times higher than the solar abundance, due to the smaller day to night contrast and the smoother transition between them.
To quantify the reliability of the solutions given by \taurex, we plotted the absorption contribution plots of the solutions for the \gcm case and the differences in ppm between the generated and the retrieved spectrum in Fig \ref{fig: contribution-gcm}. We highlight here in which wavelength bands the fit is good or not. Moreover, the calculated \redchi\ which is below 1 for each simulation, demonstrates a high significance level of the solution given by \taurex (see Fig \ref{fig: comparison plot summarize}).

\subsubsection{Thermal dissociation of \hh\ and \hho}
\label{h2ovar-h2var}

In Fig \ref{fig: comparison plot summarize} we give the \redchi\ values for each simulations. We see that TauREx finds non-degenerate solution for every simulation, although they are not all statistically significant, i.e. \redchi$\,\gg\,1$ in \Tp and \gcm cases. In the previous cases where we have a constant composition in the entire atmosphere and in the case where only water dissociates in the day side, \taurex finds a consistent solution, even though it does not necessarily converge towards the correct input parameters. 
On the contrary, when H$_2$ dissociation is taken into account, TauREx cannot find a suitable solution to explain the atmospheric spectrum. As shown in the lower left panel of \fig{fig: comparison plot summarize}, the best reduced $\chi^2$ for in \Tp and \gcm models with H$_2$ dissociation are respectively around 30 and 8. The $\chi^2$ test can give us a possible signal that in the planet we are taking into account it occurs H$_2$ dissociation.
However, \taurex manages to find statistically significant solution in the \Tm case, but we keep in mind that this simulation is a conservative choice and we think that the truth remains in between \Tm and \Tp simulations.

We note that the \taurex retrieval code is not designed to take into account the thermal dissociation of \hh. When \hh\ dissociation occurs, the \he/\hh\ ratio changes as a function of pressure and temperature. The mean atmospheric weight can also be lower than 2amu in the atmospheric regions where \hh\ dissociates. In order to be consistent with all the set of simulations, we used the same configuration file for our spectral retrievals and we left the \he/\hh\ ratio constant. By doing so we could estimate whether an atmosphere where \hh\ dissociation does not occur can explain our input spectrum or not.

The \hh\ dissociation plays a major role on the spectra and the retrievals results. All the results are shown in the light yellow part of Fig \ref{fig: comparison plot summarize}.
In each simulation, the \COratio ratio retrieved is higher than the one calculated when \hh\ is constant. As \co\ abundance remains constant everywhere, its features are, then, more significant as explained in Sect. \ref{3D trans spectra} and \taurex finds a best fit model with a low water abundance and a high \co\ abundance. As a consequence, the \COratio ratio increases.

As shown in the light yellow part in Fig \ref{fig: comparison plot summarize}, when we consider \hh\ dissociation in the \gcm case \taurex finds non-degenerate solution. The temperature retrieved is a bit higher than the temperature we find when \hh\ does not dissociate. Indeed, as shown in Fig \ref{fig: Spectra_H2cst_gcm}, \hh\ dissociation does not alter the amplitude of the water features dramatically. This is because the water absorption comes from the cold night side of the planet where molecular \hh\ dominates. Thus, the temperature retrieved still corresponds to the temperature of the night side.

For the other retrieved parameters, we are consistent with the solution found when \hh\ is constant. We remember that the dissociation of \hh\ affects the transmission spectrum in mainly 2 ways: i) it increases the scale height of the day side of the atmosphere implying stronger features for species who remains in the day side such as \co. We only consider this effect in the \Tp simulation; ii) it also significantly affects the thermochemical properties of the atmosphere due to the recombination of \hh\ in the night side which redistribute the energy, heating the night side and cooling the day side \citep{Bell_2018, TK19}. We took into account both of them in the \Tm\ simulation. \hh\ dissociation results also in a decrease of $\mathrm{H}{_2}$-$\mathrm{H}{_2}$ and He-$\mathrm{H}{_2}$ collisions, hence less intense continuum absorptions. However, in the \Tp simulations, \taurex does not manage to converge to a reasonable solution. \taurex cannot explain the line distortion using a 1D atmospheric model and tends to increase the temperature far above the equilibrium temperature of the planet, reaching the upper prior range temperature, i.e. 3600K. (see Fig. \ref{fig: comparison plot summarize}).

When both \hh\ and \hho\ dissociate, it is not possible to find any 1D atmospheric model that matches with the input spectrum in any of the cases under study in this paper.

\section{Discussions}
\label{sec:discus}

Our results show that a 1D model is always able to find a statistically consistent solution -- except for the case in which \hh\ dissociation occurs -- although it does not always converge towards the correct input solution.

\citet{Caldas2019} showed that day-night temperature contrasts lead to a bias in retrieved temperatures (toward the one of the day side) even when the chemical composition is uniform. 
In this work, we extend this by showing that when we take into account H$_2$O dissociation, we have both composition and temperature biases.

Although a retrieval code can use a 1D atmospheric model to \textit{detect} the presence of a particular chemical species from an atmosphere with a complex 3D structure, \textit{measuring} its abundance is more challenging. When temperature and chemical composition vary across the limb, the 1D retrieval cannot find the correct molecular abundances: the best fit parameters can be several orders of magnitude different from the correct input ones, as shown in Fig \ref{fig: comparison plot summarize}.

Elemental ratios (such as C/O, C/N, etc.) are a key parameter determining the chemical processes taking place in planetary atmospheres \citep{Line2013, Oreshenko2017}. At high temperature, \citet{Madhusudhan_2012, Espinoza_2017} showed that the C/O ratio plays a major role in the atmospheric chemistry and non-solar values could explain observations for 6 hot Jupiters (XO-1b, CoRoT-2b, WASP-14b, WASP-19b, WASP-33b, and WASP-12b). 
However, these elemental ratios cannot be directly measured, and we have to rely on measuring the abundance of all molecules carrying the considered elements. In the case of the hot, hydrogen dominated atmospheres considered here, H$_2$O and CO are the main carriers of carbon and oxygen so that the \COratio is commonly expected to provide reasonable constraints on the C/O ratio. Our study shows that using \COratio to constrain the C/O ratio when dissociation is present is highly hazardous and that finding $[\mathrm{CO}]/[\mathrm{H_2O}] \gg 1$ should be seen as a sign of chemical heterogeneities in the atmosphere.
We note that the same diagnostic could be applied to other species such as TiO or VO by calculating the retrieved \TiOratio or \VOratio ratio.

The fact that our retrieval results are consistent for both the \gcm model and our more idealized models of WASP-121b where temperature is constant in the vertical and symmetric around the substellar point reveals that, for UHJs at least, the effect of thermal and compositional heterogeneities \textit{across} the limb dominate over the vertical ones as well as the ones \textit{along} the limb. 
Fixing an atmospheric configuration -- i.e. in presence or not of H$_2$ and/or H$_2$O dissociation, corresponding to the red, dark yellow, and light yellow areas in Fig \ref{fig: comparison plot summarize} -- the retrieved \COratio ratio for the \Tp, \Tm, and \gcm simulations are almost the same. This result shows that the idealized cases capture the most salient features of the \gcm simulation. The most important point is that the retrieved \COratio ratio reaches its minimum value when molecular dissociation does not occur, increases by a factor 10-100 when H$_2$O dissociates, and by a factor 25-1000 when H$_2$ dissociates as well, irrespective of the details of the vertical structure of equator-to-pole thermal contrasts. This large scale in the \COratio ratio retrieved are due to our idealized \Tp and \Tm simulations which represent extreme cases, thus they encompass the biases. 

A general characteristics of our results is that, in all the cases under our study, we cannot converge around the ground truth parameters. When we study a transmission spectrum extracted from a 3D atmosphere even a high signal to noise spectrum does not allow you to know the real chemical distribution of elements with a retrieval code which uses a 1D atmospheric approximation.

A second aspect is that, with a \redchi-test it is possible to understand whether there is some more complex physical phenomenon in the atmosphere under study. In our case, we see that a \redchi$\,> 1$ was due to the presence of H$_2$ dissociation in our atmosphere. From this point of view, a \redchi-test can reveal whether there is an important physical phenomenon that we are not considering.

Note that a limitation of our study concerns the chemical components of the atmosphere. We only consider two absorber species, \co\ and \hho, but we know that other species have been detected in \wasp such as TiO or VO. Plus, we also know that we are hot enough to dissociate those species \citep{Parmentier2018} which would add more complexity in the retrieval analysis.

\section{Are there hints of 3D structures in real data?}

\begin{figure}
\centering
\includegraphics[scale=0.55]{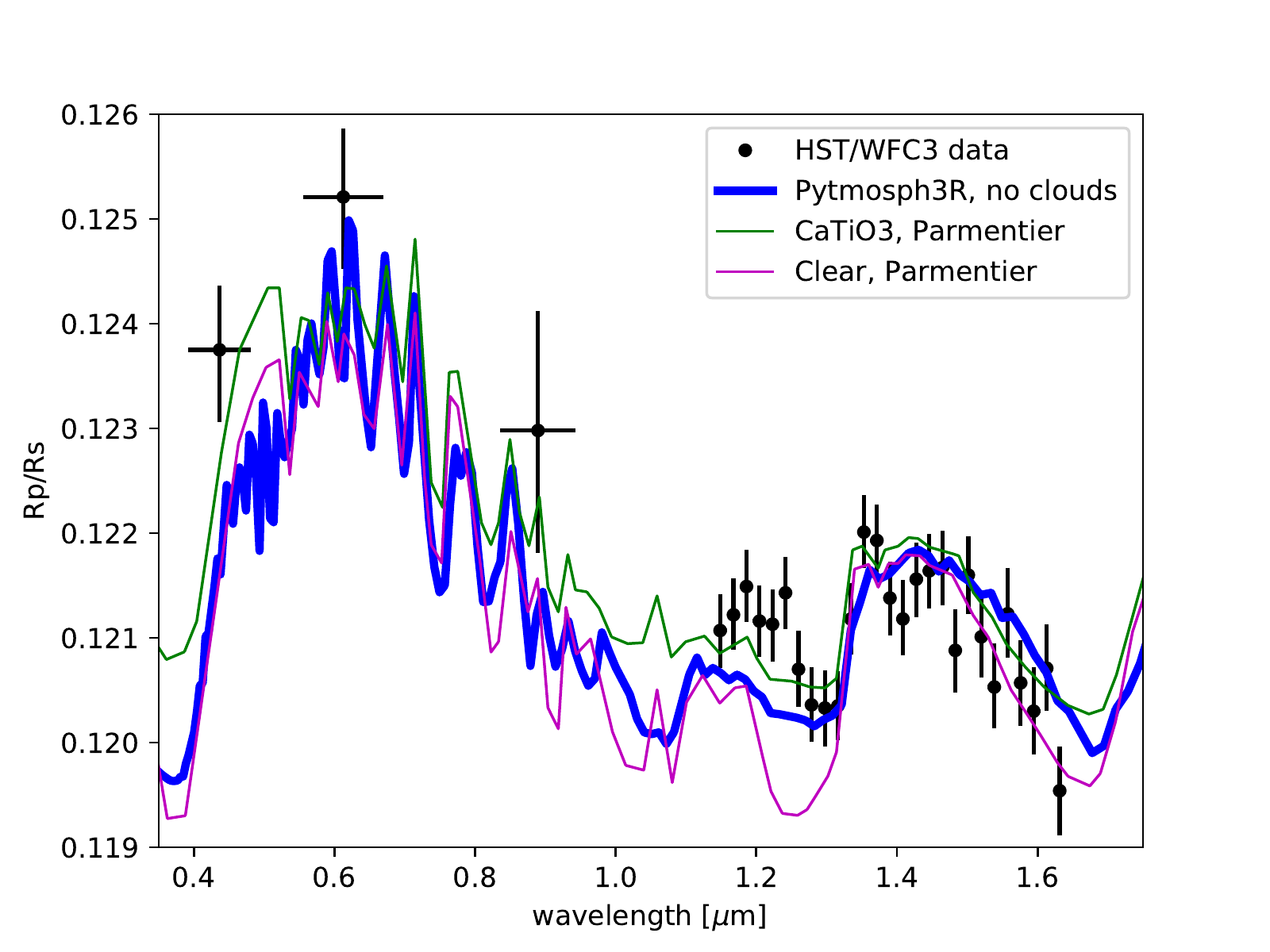}
\caption{Transmission spectra of \wasp at resolution of R = 100 for an atmosphere composed of He, \hh, \hho, \co, \tio, \vo, \na, \k and H$^{-}$ using the analytical fit for the thermal dissociation of the species in \citet{Parmentier2018} (Blue). We compared the models with HST/WFC3 data taken from \citet{Evans2016} (black dot) and with two models from \citet{Parmentier2018}, the first without clouds (magenta) and the second with CaTiO3 clouds (green).}
\label{fig:Spectra_allspecies_comp}%
\end{figure}

In \fig{fig:Spectra_allspecies_comp} we compare the spectrum of \wasp as observed with HST/WFC3 \citep{Evans2016} with two sets of synthetic transmission spectra: i) the models developed by \citet{Parmentier2018} that extract the columns at the terminator of their GCM simulation to compute the transmission spectrum, and ii) our models based on the same GCM simulation but with our more complete 3D radiative transfer framework. In addition to the radiatively active species discussed above, we account here for absoptions by TiO, VO, Na and H$^-$ using the analytical fit for the thermal dissociation of the species in \citet{Parmentier2018}. Compared to \citet{Parmentier2018}, only FeH is missing, but it is believed to be less important in this part of the spectrum since we assume a solar composition.

In order to fit HST/WFC3 data, \citet{Parmentier2018} suggests to add CaTiO$_3$ clouds in the atmosphere to increase the absorption at 1.25 $\mu$m and fill the water window region.
From our analysis, we see that considering a cloud free 3D structure can equivalently explain water amplitude in the WFC3 bandpass (thick blue line) even if our spectrum does not fit very good the data either. 3D effects can, indeed, deform the shape of the spectrum. The reason is however quite different. Instead of fitting the 1.25 $\mu$m window region, here the agreement is reached by reducing the strength of the 1.15 and 1.4 $\mu$m water bands thanks to water removal on the day side of the atmosphere. The fact that both models can have the same observed transit depth at these two wavelengths in \fig{fig:Spectra_allspecies_comp} is just due to the fact that the radius at the base of the model remains a free parameter allowing us to shift the whole spectra vertically. Note that, expect for this parameter, there is no fitting of the abundances or thermal structure involved in our approach.
We shown that because of the water dissociation, 3D effects was not negligible for atmospheres as hot as wasp-121b, even if clouds remains a reasonable assumption to fit the data. This findings suggests that both 3D heterogeneities in the transmission spectra and clouds can be combined to explain the reduction of the water feature. However, this is depending on which altitude the clouds remain and where they are in the atmosphere because even if there is clouds, the light rays could not probe them due to the 3D structure of the atmosphere. 

\section{Conclusion}

The 3D structure of the planetary atmospheres plays a major role in shaping of the transmission spectra, especially for ultra hot Jupiters. The tidally locked planets have an inflated day side due to both the high temperature and the dissociation of species which increase the scale height of the day side atmosphere.
In these atmospheres, the transmitted spectrum is affected by an extended region around the terminator line, which includes part of both the day side and the night side of the atmosphere.

The stellar light probes an atmospheric region which extends significantly toward the limb, depending on the chemical composition as a function of wavelength. If the temperature is not high enough to dissociate all the molecules, such as \co, they will remain everywhere in the atmosphere. Thus, the amplitude of their features in the transmission spectrum will be larger because they come from the hot regions if the inflated day side atmosphere. On the other hand, the molecules which dissociate more easily, such as \hho, will only remain in the cold night side of the atmosphere. Therefore, the amplitude of the features of those molecules in the transmission spectrum will be less pronounced, since they come from colder regions in the night side of the atmosphere. A 1D retrieval code, which tries to fit the spectrum with an isothermal atmosphere, cannot unravel the information of a more complex temperature distribution and, then the abundances retrieved are biased. In particular, we demonstrate that for UHJs 1D retrieval would be biased towards high [CO]/[H$_2$O] ratio, the higher the ratio the stronger the chemical heterogeneities. For instance, the \taurex retrieval code finds the planetary temperature according to the strength of the main spectral features. Since the amplitude of the spectral features is rather large because of the gas present in the hot day side of the planet, it is natural for a retrieval code to converge towards hotter temperatures. In case of H$_2$O dissociation, the only relevant gas present in the hot day-side is the CO. The water absorption is due to the water present in the cold night side of the planet. Therefore, it is reasonable to retrieve a lower temperature in the case in which water dissociates in the day side.

We demonstrated that a three-dimensional atmospheric structure induces spectral distortions impossible to explain with a 1D retrieval. We also demonstrate that the water features can be strongly reduced by the presence of molecular heterogeneities due to a large day/night temperature contrast. For instance, we show that a 3D geometry can explain some of the features observed in the HST/WFC3 wavelength range of \wasp, even in absence of hazes or clouds, thanks to the dissociation of water in the day side of the atmosphere. However, our works do not exclude the presence of clouds in UHJ atmospheres, the 3D effects are a complementary contribution to fill the water windows in the data. Thanks to our results, we think that \wasp would be so a good target for future observations. We propose that 3D structural effects should already be consider when we study hot and ultra hot Jupiter to avoid erroneous conclusions. Moreover, future space missions such as JWST \citep{Beichman2014} and ARIEL \citep{Tinetti2018} will probe a very large range in wavelength (from 0.6 to 28 $\mu$m for JWST) and, then 3D effects will be even more evident in other parts of the exoplanetary spectra which has not been observed yet.

Finally, after the analysis of an exoplanetary atmosphere with a 1D retrieval tool, the \redchi-test on a 1D retrieval can raise a warning about an important effect that the 1D model cannot consider, such as effects induced by the 3-dimensional structure of the exoplanetary atmosphere. In such cases, a parametrized 2D retrieval approach may be warranted. Whether there is enough information in the spectrum to actually constrain such a 2D approach remains to be demonstrated.

\begin{acknowledgements}
This project has received funding from the European Research Council (ERC) under the European Union's Horizon 2020 research and innovation programme (grant agreement n$^\circ$679030/WHIPLASH).
\end{acknowledgements}

\bibliographystyle{aa}
\bibliography{biblio_wasp121b}

\begin{thebibliography}{51}
\expandafter\ifx\csname natexlab\endcsname\relax\def\natexlab#1{#1}\fi

\bibitem[{Arcangeli {et~al.}(2018)Arcangeli, D{\'{e}}sert, Line, Bean,
  Parmentier, Stevenson, Kreidberg, Fortney, Mansfield, \&
  Showman}]{Arcangeli_2018}
Arcangeli, J., D{\'{e}}sert, J.-M., Line, M.~R., {et~al.} 2018, The
  Astrophysical Journal, 855, L30

\bibitem[{{Arcangeli} {et~al.}(2019){Arcangeli}, {D{\'e}sert}, {Parmentier},
  {Stevenson}, {Bean}, {Line}, {Kreidberg}, {Fortney}, \&
  {Showman}}]{Arcangeli2019}
{Arcangeli}, J., {D{\'e}sert}, J.-M., {Parmentier}, V., {et~al.} 2019, \aap,
  625, A136

\bibitem[{Barton {et~al.}(2014)Barton, Chiu, Golpayegani, Yurchenko, Tennyson,
  Frohman, \& Bernath}]{Barton2014}
Barton, E.~J., Chiu, C., Golpayegani, S., {et~al.} 2014, \mnras, 442, 1821

\bibitem[{Barton {et~al.}(2013)Barton, Yurchenko, \& Tennyson}]{Barton2013}
Barton, E.~J., Yurchenko, S.~N., \& Tennyson, J. 2013, \mnras, 434, 1469

\bibitem[{Beichman {et~al.}(2014)Beichman, Benneke, \& Knutson}]{Beichman2014}
Beichman, C., Benneke, B., \& Knutson, H. e.~a. 2014, \pasp, 126, 1134

\bibitem[{{Bell} \& {Cowan}(2018)}]{BC18}
{Bell}, T.~J. \& {Cowan}, N.~B. 2018, \apjl, 857, L20

\bibitem[{Bell \& Cowan(2018)}]{Bell_2018}
Bell, T.~J. \& Cowan, N.~B. 2018, The Astrophysical Journal, 857, L20

\bibitem[{Caldas {et~al.}(2019)Caldas, Leconte, \& Selsis}]{Caldas2019}
Caldas, A., Leconte, J., \& Selsis, F. e.~a. 2019, \aap, 623, A161

\bibitem[{Changeat {et~al.}(2019)Changeat, Edwards, Waldmann, \&
  Tinetti}]{Changeat_2019}
Changeat, Q., Edwards, B., Waldmann, I.~P., \& Tinetti, G. 2019, The
  Astrophysical Journal, 886, 39

\bibitem[{{Cowan} {et~al.}(2015){Cowan}, {Greene}, {Angerhausen}, {Batalha},
  {Clampin}, {Col{\'o}n}, {Crossfield}, {Fortney}, {Gaudi}, {Harrington},
  {Iro}, {Lillie}, {Linsky}, {Lopez-Morales}, {Mandell}, \&
  {Stevenson}}]{2015PASP..127..311C}
{Cowan}, N.~B., {Greene}, T., {Angerhausen}, D., {et~al.} 2015, \pasp, 127, 311

\bibitem[{Delrez {et~al.}(2016)Delrez, Santerne, \& Almenara}]{Delrez2016}
Delrez, L., Santerne, A., \& Almenara, J.-M. e.~a. 2016, \mnras, 458, 4025

\bibitem[{Espinoza {et~al.}(2017)Espinoza, Fortney, Miguel, Thorngren, \&
  Murray-Clay}]{Espinoza_2017}
Espinoza, N., Fortney, J.~J., Miguel, Y., Thorngren, D., \& Murray-Clay, R.
  2017, The Astrophysical Journal, 838, L9

\bibitem[{Evans {et~al.}(2017)Evans, Sing, \& Kataria}]{Evans2017}
Evans, T.~M., Sing, D.~K., \& Kataria, T. e.~a. 2017, \nat, 548, 58

\bibitem[{Evans {et~al.}(2016)Evans, Sing, \& Wakeford}]{Evans2016}
Evans, T.~M., Sing, D.~K., \& Wakeford, H. R. e.~a. 2016, \apjl, 822, L4

\bibitem[{Fu \& Liou(1992)}]{Fu1992}
Fu, Q. \& Liou, K.~N. 1992, Journal of Atmospheric Sciences, 49, 2139

\bibitem[{{Goody} \& {Yung}(1989)}]{Goody1989}
{Goody}, R.~M. \& {Yung}, Y.~L. 1989, {Atmospheric radiation : theoretical
  basis}

\bibitem[{Gordon {et~al.}(2010)Gordon, Kassi, Campargue, \& Toon}]{Gordon2010}
Gordon, I.~E., Kassi, S., Campargue, A., \& Toon, G.~C. 2010, in 65th
  International Symposium On Molecular Spectroscopy, WF03

\bibitem[{Gordon {et~al.}(2013)Gordon, Rothman, \& Li}]{Gordon2013}
Gordon, I.~E., Rothman, L.~S., \& Li, G. 2013, in 68th International Symposium
  on Molecular Spectroscopy, ERE03

\bibitem[{Greene {et~al.}(2016)Greene, Line, Montero, Fortney, Lustig-Yaeger,
  \& Luther}]{Greene_2016}
Greene, T.~P., Line, M.~R., Montero, C., {et~al.} 2016, The Astrophysical
  Journal, 817, 17

\bibitem[{Haynes {et~al.}(2015)Haynes, Mandell, \& Madhusudhan}]{Haynes2015}
Haynes, K., Mandell, A.~M., \& Madhusudhan, N. e.~a. 2015, \apj, 806, 146

\bibitem[{{Kataria} {et~al.}(2015){Kataria}, {Showman}, {Fortney}, {Stevenson},
  {Line}, {Kreidberg}, {Bean}, \& {D{\'e}sert}}]{Kataria2015}
{Kataria}, T., {Showman}, A.~P., {Fortney}, J.~J., {et~al.} 2015, \apj, 801, 86

\bibitem[{{Kataria} {et~al.}(2013){Kataria}, {Showman}, {Lewis}, {Fortney},
  {Marley}, \& {Freedman}}]{Kataria2013}
{Kataria}, T., {Showman}, A.~P., {Lewis}, N.~K., {et~al.} 2013, \apj, 767, 76

\bibitem[{{Kreidberg} {et~al.}(2018){Kreidberg}, {Line}, {Parmentier},
  {Stevenson}, {Louden}, {Bonnefoy}, {Faherty}, {Henry}, {Williamson}, \&
  {Stassun}}]{Kreidberg2018}
{Kreidberg}, L., {Line}, M.~R., {Parmentier}, V., {et~al.} 2018, \aj, 156, 17

\bibitem[{Kreidberg {et~al.}(2018)Kreidberg, Line, Parmentier, Stevenson,
  Louden, Bonnefoy, Faherty, Henry, Williamson, Stassun, Beatty, Bean, Fortney,
  Showman, D{\'{e}}sert, \& Arcangeli}]{Kreidberg_2018}
Kreidberg, L., Line, M.~R., Parmentier, V., {et~al.} 2018, The Astronomical
  Journal, 156, 17

\bibitem[{Lenzuni {et~al.}(1991)Lenzuni, Chernoff, \& Salpeter}]{Lenzuni1991}
Lenzuni, P., Chernoff, D.~F., \& Salpeter, E.~E. 1991, \apjs, 76, 759

\bibitem[{{Lewis} {et~al.}(2017){Lewis}, {Parmentier}, {Kataria}, {de Wit},
  {Showman}, {Fortney}, \& {Marley}}]{Lewis2017}
{Lewis}, N.~K., {Parmentier}, V., {Kataria}, T., {et~al.} 2017, ArXiv e-prints:
  1706.00466

\bibitem[{Line {et~al.}(2013)Line, Wolf, \& Zhang}]{Line2013}
Line, M.~R., Wolf, A.~S., \& Zhang, X. e.~a. 2013, \apj, 775, 137

\bibitem[{Lodders \& Fegley(2002)}]{Lodders2002}
Lodders, K. \& Fegley, B. 2002, \icarus, 155, 393

\bibitem[{Madhusudhan(2012)}]{Madhusudhan_2012}
Madhusudhan, N. 2012, The Astrophysical Journal, 758, 36

\bibitem[{{Marley} \& {McKay}(1999)}]{Marley1999}
{Marley}, M.~S. \& {McKay}, C.~P. 1999, \icarus, 138, 268

\bibitem[{Marley {et~al.}(2017)Marley, Saumon, \& Fortney}]{Marley2017}
Marley, M.~S., Saumon, D., \& Fortney, J. J. e.~a. 2017, in American
  Astronomical Society Meeting Abstracts, Vol. 230, American Astronomical
  Society Meeting Abstracts \#230, 315.07

\bibitem[{Oreshenko {et~al.}(2017)Oreshenko, Lavie, \& Grimm}]{Oreshenko2017}
Oreshenko, M., Lavie, B., \& Grimm, S. L. e.~a. 2017, \apjl, 847, L3

\bibitem[{Parmentier {et~al.}(2016)Parmentier, Fortney, \&
  Showman}]{Parmentier2016}
Parmentier, V., Fortney, J.~J., \& Showman, A. P. e.~a. 2016, \apj, 828, 22

\bibitem[{Parmentier {et~al.}(2018)Parmentier, Line, \& Bean}]{Parmentier2018}
Parmentier, V., Line, M.~R., \& Bean, J. L. e.~a. 2018, \aap, 617, A110

\bibitem[{{Parmentier} {et~al.}(2013){Parmentier}, {Showman}, \&
  {Lian}}]{Parmentier2013}
{Parmentier}, V., {Showman}, A.~P., \& {Lian}, Y. 2013, \aap, 558, A91

\bibitem[{Rothman {et~al.}(2009)Rothman, Gordon, \& Barbe}]{Rothman2009}
Rothman, L.~S., Gordon, I.~E., \& Barbe, A. e.~a. 2009, \jqsrt, 110, 533

\bibitem[{Sheppard {et~al.}(2017)Sheppard, Mandell, \& Tamburo}]{Sheppard2017}
Sheppard, K.~B., Mandell, A.~M., \& Tamburo, P. e.~a. 2017, \apjl, 850, L32

\bibitem[{{Showman} {et~al.}(2009){Showman}, {Fortney}, {Lian}, {Marley},
  {Freedman}, {Knutson}, \& {Charbonneau}}]{Showman2009}
{Showman}, A.~P., {Fortney}, J.~J., {Lian}, Y., {et~al.} 2009, \apj, 699, 564

\bibitem[{Showman {et~al.}(2015)Showman, Lewis, \& Fortney}]{Showman2015}
Showman, A.~P., Lewis, N.~K., \& Fortney, J.~J. 2015, \apj, 801, 95

\bibitem[{Stevenson {et~al.}(2016)Stevenson, Lewis, \& Jacob
  L.~Bean}]{Stevenson_2016}
Stevenson, K.~B., Lewis, N.~K., \& Jacob L.~Bean, e.~a. 2016, Publications of
  the Astronomical Society of the Pacific, 128, 094401

\bibitem[{Sudarsky {et~al.}(2000)Sudarsky, Burrows, \& Pinto}]{Sudarsky_2000}
Sudarsky, D., Burrows, A., \& Pinto, P. 2000, The Astrophysical Journal, 538,
  885

\bibitem[{{Tan} \& {Komacek}(2019)}]{TK19}
{Tan}, X. \& {Komacek}, T.~D. 2019, arXiv e-prints, arXiv:1910.01622

\bibitem[{Tennyson \& Yurchenko(2012)}]{Tennyson2012}
Tennyson, J. \& Yurchenko, S.~N. 2012, \mnras, 425, 21

\bibitem[{Tinetti {et~al.}(2018)Tinetti, Drossart, \& Eccleston}]{Tinetti2018}
Tinetti, G., Drossart, P., \& Eccleston, P. e.~a. 2018, Experimental Astronomy,
  46, 135

\bibitem[{Visscher {et~al.}(2006)Visscher, Lodders, \& Fegley}]{Visscher2006}
Visscher, C., Lodders, K., \& Fegley, Jr., B. 2006, \apj, 648, 1181

\bibitem[{Visscher {et~al.}(2010)Visscher, Lodders, \& Fegley}]{Visscher2010}
Visscher, C., Lodders, K., \& Fegley, Jr., B. 2010, \apj, 716, 1060

\bibitem[{Waldmann {et~al.}(2015{\natexlab{a}})Waldmann, Rocchetto, \&
  Tinetti}]{Waldmann2015a}
Waldmann, I.~P., Rocchetto, M., \& Tinetti, G. e.~a. 2015{\natexlab{a}}, \apj,
  813, 13

\bibitem[{Waldmann {et~al.}(2015{\natexlab{b}})Waldmann, Tinetti, \&
  Rocchetto}]{Waldmann2015}
Waldmann, I.~P., Tinetti, G., \& Rocchetto, M. e.~a. 2015{\natexlab{b}}, \apj,
  802, 107

\bibitem[{Wright {et~al.}(2012)Wright, Marcy, \& Howard}]{Wright2012}
Wright, J.~T., Marcy, G.~W., \& Howard, A. W. e.~a. 2012, \apj, 753, 160

\bibitem[{Yurchenko {et~al.}(2011)Yurchenko, Barber, \&
  Tennyson}]{Yurchenko2011}
Yurchenko, S.~N., Barber, R.~J., \& Tennyson, J. 2011, \mnras, 413, 1828

\bibitem[{Yurchenko {et~al.}(2014)Yurchenko, Tennyson, \&
  Bailey}]{Yurchenko2014}
Yurchenko, S.~N., Tennyson, J., \& Bailey, J. e.~a. 2014, Proceedings of the
  National Academy of Science, 111, 9379

\end{thebibliography}
\end{document}